\documentclass[a4paper,fleqn,usenatbib,useAMS]{mnras}

\usepackage{graphicx}	
\usepackage{amsmath}	
\usepackage{amssymb}	
\usepackage{multicol}        
\usepackage{bm}		
\usepackage{pdflscape}	

\usepackage{newtxtext,newtxmath}

\usepackage[T1]{fontenc}
\usepackage{ae,aecompl}

\title[DO-CRIME]{DO-CRIME: Dynamic On-sky Covariance Random Interaction Matrix Evaluation, a novel method for calibrating adaptive optics systems.}

\author[O. Lai et al.]{
Olivier Lai$^{1}$\thanks{E-mail: olivier.lai\@oca.eu},
Mark Chun$^{2}$,
Ryan Dungee$^{2}$,
Jessica Lu$^{3}$,
and Marcel Carbillet$^{1}$
\\
$^{1}$Universit\'{e} C\^{o}te d'Azur, Observatoire de la C\^{o}te d'Azur, CNRS, 
Laboratoire Lagrange, Bd de l'Observatoire, CS 34229, 06304 Nice cedex 4, France \\
$^{2}$Institute for Astronomy, University of Hawaii, 640 North A`oh\={o}k\={u} Place, Hilo 96720, USA\\
$^{3}$Department of Astronomy, University of California at Berkeley, Berkeley CA 94720, USA\\
}

\date{Accepted XXX. Received YYY; in original form ZZZ}

\pubyear{2020}

\begin{document}
\label{firstpage}
\pagerange{\pageref{firstpage}--\pageref{lastpage}}
\maketitle

\begin{abstract}
Adaptive optics systems require a calibration procedure to operate, whether in closed loop 
or even more importantly in forward control. This calibration usually takes the form of an interaction 
matrix and is a measure of the response on the 
wavefront sensor to wavefront corrector stimulus. If this matrix is sufficiently well conditioned, it can 
be inverted to produce a control matrix, which allows to compute the optimal commands to apply to the wavefront 
corrector for a given wavefront sensor measurement vector. Interaction matrices are usually measured by 
means of an artificial source at the entrance focus of the adaptive optics system; however, adaptive secondary 
mirrors on Cassegrain telescopes offer no such focus and the measurement of their interaction matrices becomes 
more challenging and needs to be done on-sky using a natural star. The most common method is to generate a 
theoretical or simulated interaction matrix and adjust it parametrically (for example, decenter, magnification, 
rotation) using on-sky measurements.  We propose a novel  method of measuring on-sky interaction matrices 
{\em ab initio} from the telemetry stream of the AO system using random patterns on the deformable mirror with 
diagonal commands covariance matrices. The approach, being developed for the adaptive secondary mirror 
upgrade for the imaka wide-field AO system on the UH2.2m telescope project, is shown to work on-sky using the 
current imaka testbed. 
\end{abstract}




\section{Introduction}
The foundational principle of adaptive optics (AO) is to measure a wavefront disturbance using a wavefront 
sensor (WFS) and to correct the measured aberrations in real time using a wavefront corrector, usually a 
deformable mirror (DM) but potentially, any phase inducing device (LCD, lenses misalignment, etc.) 
could be used \citep{ref:babcock53}. A computer is used to translate the wavefront sensor measurements into DM commands 
in real time. In effect the real time computer needs to solve a set of linear equations relating the measurement 
vector $\mathbf{m}$ to the commands vector $\mathbf{c}$. 
\begin{equation}
\label{eq:linearsys}
\mathbf{c} = A \cdot \mathbf{m}
\end{equation}
where $A$ is a matrix usually called the control matrix \citep{ref:tyson91, ref:roddier99}. 
At optical and infrared wavelengths, it is usually not possible to measure the phase of a light wave directly, 
so an optical device has to be used to transform a function of the 
phase into a measurable intensity, which is why a non-trivial set of linear equations is needed to reconstruct the 
wavefront. For example, interferometric sensors transform the sine of the phase into intensity (usually leading 
to a limited linear range and a 2 $\pi$ ambiguity); Shack-Hartman and pyramid
wavefront sensors  transform the gradient 
of the phase into an intensity centroid measurement and curvature sensors transform the laplacian of the phase 
into an intensity difference measurement. The advantage of the latter comes from their use with bimorph mirrors, 
which produce a constant curvature over their electrodes. In theory, the one-to-one correspondance 
between the measurements and the correction is equivalent to a perfectly diagonal matrix $A$. In practice, 
there is sufficient cross-talk in curvature systems to warrant a wavefront 
reconstructor such as the one described by equation~\ref{eq:linearsys} (i.e. with non-diagonal terms in $A$). 
All modern AO systems use some variant of 
Equation~\ref{eq:linearsys} in their control system and optimally mapping the actuator to the measurements 
is the topic of numerous studies.

The most common way to obtain the control matrix $A$ is to measure the so-called interaction matrix which 
we will call $D$. It is constructed by exciting each actuator or degree of freedom of the wavefront corrector 
sequentially and recording the the wavefront sensor measurements such that:
\begin{equation}
\label{eq:fundamat}
\mathbf{m} = D \cdot \mathbf{c} 
\end{equation}
We might be tempted to write that $D = \mathbf{m} \cdot \mathbf{c}^{-1} $, but this equation doesn't 
capture the process of acquiring the interaction matrix by 
sequentially exciting linear combinations of commands or modes. Furthermore the inverse of a vector is 
not a well defined concept; although in the above equation, it can be thought of as 
$\mathbf{c}^{-1} = \mathbf{c}^T/|\mathbf{c}|^2$. 

Instead, we concatenate each sequential command to write a matrix of commands $C$, where each column
corresponds to an actuation of the deformable mirror. For example, sequentially 
exciting each actuator would make $C$ a diagonal matrix. Multiplying $D$ by $C$ produces a matrix of 
measurements $M$, in which each column corresponds to the response vector associated with command vector 
(the $n^{th}$ column of $C$, $\mathbf{c}_{i,n}$). We can then write:
\begin{eqnarray}
M & = & D \cdot C\\
D & = & M \cdot C^{-1} 
\end{eqnarray}

At this point we note that nothing predetermines the matrix of commands $C$ to be diagonal, and it has been proposed
that an optimal method of measuring interaction matrices
consists in making $C$ a Hadamard matrix to generate the strongest signal, the strongest diversity of measurements 
on the wavefront sensor \citep{ref:kasper04, ref:meimon15}. However, since the command matrix $C$ matrix has to 
be inverted, there are advantages to it being diagonal (in which case it is simply $C^{-1} = C^T/|C|^2$). This method 
is what we will call the poke matrix in the following.  

If $D$ is well conditioned then a general solution for A is given by the Moore-Penrose inverse $D^+$:
\begin{equation}
A = D^+ = (D^T \cdot D)^{-1} \cdot D^T
\end{equation}
However, in practice, $D$ is more often inverted with Singular Value Decomposition (SVD), which 
allows to filter eigenmodes of $D$ which have particularly low eigenvalues, e.g \citep{ref:gendron94, ref:lai00}; 
physically these are modes that are poorly sensed by the system, they produce 
a small measurement vector given a normalised stimulus. 
If not filtered, such modes are very sensitive to noise in the measurement vector and are amplified. 
Singular value decomposition is a generalisation of eigenmodes for non-square matrices using polar 
decomposition. We can write the interaction matrix $D$ as the product of three matrices $U, \Sigma$ 
and $V^*$:
\begin{equation}
D=U \cdot \Sigma \cdot V^*
\end{equation}
If $D$ is an $m \times n$ matrix (measurement vector has $m$ elements and 
DM has $n$ actuators), then $U$ is an $m \times m$ unitary matrix, $\Sigma$ is an $m \times n$ 
rectangular diagonal matrix with non-negative numbers on the diagonal and $V$ is also a unitary matrix, of dimension $n \times n$. 
As $D$ is a real matrix, $U$ and $V$are real orthonormal matrices. The diagonal entries 
$\sigma _{i}=\Sigma _{ii}$ are known as the singular values of $D$. The number of non-zero singular 
values is equal to the number of independent degrees of freedom of the AO system. 
The pseudo-inverse of D is then given by:
\begin{equation}
\label{eq:svd}
D^+ = V \cdot \Sigma^+ \cdot U^*
\end{equation}

where $\Sigma^+$ is the pseudo-inverse of $\Sigma$ which is formed by replacing every non-zero 
diagonal entry by its reciprocal and transposing the resulting matrix. If values of $\Sigma_{ii}$ are very small, 
the corresponding value $\Sigma^+_{ii}$ can be set to zero, thereby effectively filtering this particular mode, or 
degree of freedom from the reconstruction. We call the number of controlled modes $\Xi$, so it is a parameter in 
equation~\ref{eq:svd}, $D^+ (\Xi) = V \cdot \Sigma^+ (\Xi) \cdot U^*$. Different AO geometries have 
such modes, called invisible modes or the null space of the control matrix: waffle mode in the case of 
Shack-Hartman wavefront sensors in a Fried geometry, or piston for curvature sensors.

Measuring the interaction matrix D usually requires an artificial source at the input focus of the AO system, 
which allows to record the wavefront sensor measurement vector while the wavefront corrector is actuated. 
However with the development of Adaptive Secondary Mirrors (ASM), it is not always possible to introduce 
a calibration source due to the lack of an intermediate or entrance focus. This is true for Cassegrain telescopes, 
but even for Gregorian telescopes, which do have an intermediate focus before reflection on the secondary 
mirror, this focus is not always easily accessible \citep{ref:pieralli08, ref:heritier18}. 
We also note that using an artificial source may not reflect 
the state of the system when it is in use, for example Shack-Hartman spots which would be diffraction limited 
with a calibration source may be blurred by atmospheric turbulence, or the telescope pupil may not match the 
AO system pupil exactly leading to a different illumination on the lenslet array. Finally, measuring an interaction 
matrix by "poking" each actuator and waiting for the system to settle will hide dynamical effects, such as 
dynamical coupling  or variable time lag of actuators (which acts as a gain in closed loop). 
We therefore suggest that measuring interaction matrices 
on-sky and in the same conditions as during operation of the AO system has advantages beyond their use with ASMs.

Methods to generate interaction matrices when no intermediate focus is available fall into two major categories:
Methods developed mostly for the ESO Adaptive Optics Facility (AOF) are what could be described as non-invasive as 
they generate a simulated or synthetic interaction matrix using models for the influence functions and the wavefront 
sensors and then use sky data to adjust alignment parameters such as the pupil translation, magnification and rotation 
on the wavefront sensor and/or the deformable mirror. Such methods have been investigated by \cite{ref:bechet11, 
ref:bechet12}, for the VLT AOF but also in light of the future ELT.  The parametric fitting used on the AOF 
is described by \cite{ref:kolb12}  and \cite{ref:neichel12}  studied parametric adjustment of the misregistration 
for the SAXO AO system in the SPHERE high contrast  instrument, the GeMS MCAO system and the HOMER  
wide field AO testbench. The parametric adjustment of the LBT pyramid wavefront misregistration has been 
studied by \cite{ref:heritier18, ref:heritier19}.

Another method to improve synthetic or simulated interaction matrices on-sky is more invasive, as it requires modulation
of the deformable mirror (usually one or a few modes) and synchronous detection of these mode on the wavefront sensor.
These methods have mostly been developed and used at LBT \citep{ref:wildi04, ref:oberti06, ref:esposito06}. The relative 
merits of the invasive and non-invasive methods have been studied by \cite{ref:oberti06} and more recently by 
\cite{ref:heritier17} who do not reach a definite conclusion, each having advantages and drawbacks: the non-invasive 
approach obviously has no impact on observations, is fast and produces arbitrarily large SNR, but depends on complex models
of the wavefront sensor, which may be an issue for the ELT, while the invasive method works well for small misregistrations but
does have a (small) impact on observations and only a few modes can be measured at any time, so it is a slower process. Our method 
is slightly different since, although invasive, it is quite fast ($\sim$10 seconds, depending on 
the frame rate) to measure the entire interaction matrix, but unlike non-invasive methods, it is completely model
 independent, irrsepective of the scale of the misregistration.

Although the invasive and non-invasive methods work very well and are used extensively in practice, 
they have intrinsic limitations. For instance, the number of parameters to fit can 
become large if these are not global (e.g. pupil centering, magnification, rotation), and parameters can only fit 
known effects (see our example of malfunctioning electrode 10, Section~\ref{sec:electrode10} ).  Synthetic 
matrices assume perfect lenslet arrays, but small defects (in focal length or pitch) can go unnoticed in parametric fits. 
The dynamic response of the deformable mirror may not be identical for every actuator, especially as a function of temperature 
as seen on magnetic actuators deformable mirrors, \citep{ref:woillez19}, effectively acting as a 
differential gain on specific actuators. Edge subapertures also require specific care; their partial illumination 
leads to a lower signal to noise ratio, which can be mitigated by appropriate weighting. If the pupil moves with 
respect to the lenslet array during observations (for example due to flexures or pupil nutation), these weights need 
to be readjusted. This is especially critical if the centroiding algorithm uses non-linear
filters such as thresholding, or if the guide source is extended. Finally, in the case of GLAO which is of particular interest to us, 
wavefront sensors can be far off-axis, leading to pupil distortions that can be hard to parametrize, e.g. if the illumination on the 
wavefront sensor becomes elliptical or vignetted or if the actuator pattern is distorted with respect to the lenslet array. 

Therefore we suggest that measuring interaction matrices on-sky {\em ab-initio} provides a more accurate 
measurement of the state of the AO system, without any assumptions or theoretical models. 
Interaction matrices contain all the information about the system alignment and gain response and we 
suggest that measuring them regularly is a useful diagnostic tool and may improve the quality of the correction,
especially when very high performance is required. 
In cases where the deformable mirror rotates with respect to the 
lenslet array (e.g. adaptive secondary mirror with WFS at Nasmyth  or Coud\'e focus), our method can be used to 
generate on-the-fly interaction matrices or to control a pupil derotator to stabilize the WFS-DM alignement. 
The amount of on-sky time required for non-science calibration observations is low for this approach, and no 
modification is needed (e.g. introducing an artificial source), with the telescope and AO system at the same 
inclination and attitude (and hence flexures) as the science observations.

\begin{figure*}
	\includegraphics[width=17cm]{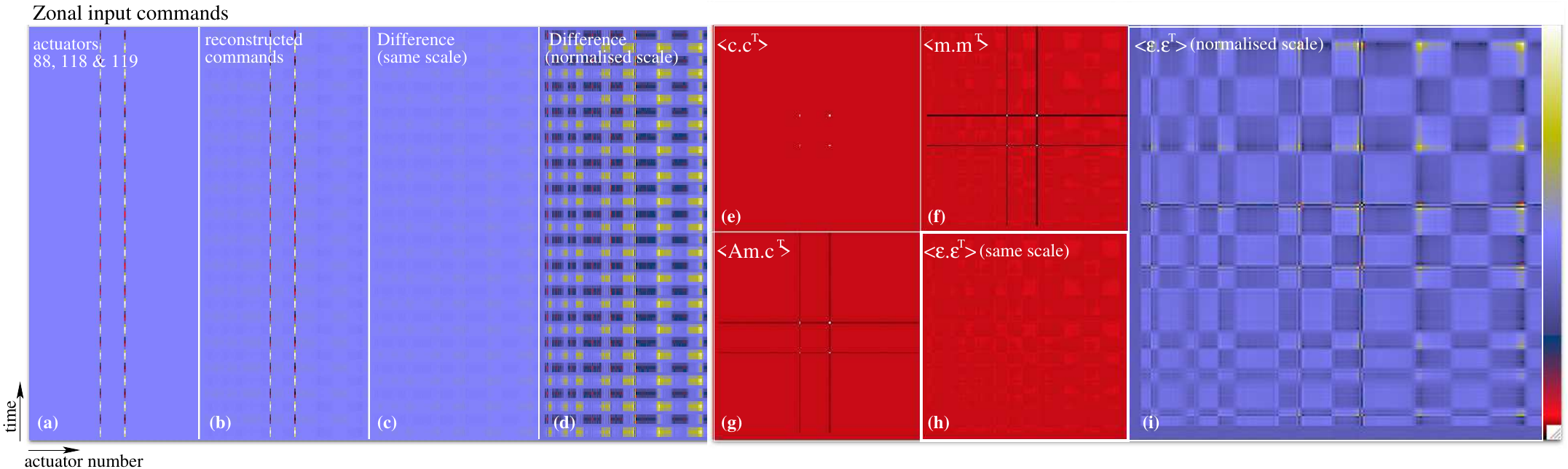}
    \caption{Zonal demonstration of the control matrix reconstruction fidelity simulation tool. Actuators 88, 118 and 119 are 
    excited sinusoidally (a), providing the input $\mathbf{c}_{a}$. Wavefront sensor measurements $\mathbf{m}_{a}$ are 
    recorded and multiplied by the control matrix $A$ (b) and the difference 
    $\mathbf{\epsilon}_{a}$ (with the same color scale) is shown in (c); (d) shows this difference but 
    stretched over the entire color scale. Panel (e) shows the covariance matrix of the commands, (f) shows the reconstructed 
    commands, (g) is the cross term of equation~\ref{eq:covepsilon} and (h) is the residual covariance 
    matrix $<\mathbf{\epsilon} \cdot \mathbf{\epsilon}^T >$. Panel (i) shows the residual covariance matrix stretched over
    the entire color scale. The trace of this matrix is the residual variance.} 
    \label{fig:modulationsim1}
\end{figure*}

\section{Method}
In closed loop, a simple integrator control loop follows the following discrete time step equation:
\begin{equation}
\mathbf{c}_{t+1} = \mathbf{c}_t + g A \cdot \mathbf{m}_t
\end{equation}
Where $g$ is a scalar loop gain (usually $<1$). Therefore, we can re-arrange and write: 
\begin{equation}
\label{eq:controlmat1}
 A \cdot \mathbf{m}_t = \frac{1}{g} (\mathbf{c}_{t+1} - \mathbf{c}_{t} )
\end{equation}
We alleviate the notation by writing $\Delta \mathbf{c}_t = \mathbf{c}_{t+1} - \mathbf{c}_t$. By multiplying 
both sides by $\mathbf{m}_{t}^{T}$ and rearranging:
\begin{equation}
 \label{eq:commat}
 A = \frac{1}{g} < \Delta \mathbf{c}_t \cdot \mathbf{m}_{t}^{T} > \cdot < (\mathbf{m}_t \cdot  \mathbf{m}_{t}^{T} ) > ^{-1}
\end{equation}
where $< >$ signfies the time average. Thus $< (\mathbf{m}_t \cdot  \mathbf{m}_{t}^{T} ) >$ is the measurements 
covariance matrix. Conversely from equations~\ref{eq:fundamat} and~\ref{eq:controlmat1}, we can also write:
\begin{equation}
\label{eq:discreteloop}
g \mathbf{m}_t  = D \Delta \mathbf{c}_t 
\end{equation}
Multuplying both sides by $ \Delta \mathbf{c}_t^T$, taking the time average and rearranging, we obtain: 
\begin{equation}
\label{eq:intmat}
D =  g < \mathbf{m}_t \cdot   \Delta \mathbf{c}_t^T > \cdot <  \Delta \mathbf{c}_t \cdot  \Delta \mathbf{c}_t^T>^{-1}
\end{equation}
$< (\Delta \mathbf{c}_t \cdot  \Delta \mathbf{c}_{t}^{T} ) >$ is the incremental commands 
 covariance matrix. The control loop bandwidth of AO systems is usually higher than the 
 atmospheric coherence time, so for atmospheric 
disturbances, the variance of the differences of subsequent commands, $\sigma_{\Delta \mathbf{c}}^2$, is small. 

Our method consists in applying a random (but known) vector $\mathbf{c}_{\xi}(t)$ of commands and recording the 
measurements. This can be done in closed loop, but if the control matrix is not known on first pass, this can 
also be done in open loop, as long as the signal remains within the linearity range of the wavefront sensor 
(this can be an issue for curvature or unmodulated pyramid sensing or Shack-Hartman using quad-cells). 
On sky in open loop, equation~\ref{eq:fundamat} can be used to write:
\begin{equation}
\label{eq:discreteopen}
\mathbf{m}_{\xi}(t) + \mathbf{m}_a(t) = D \cdot \mathbf{c}_{\xi}(t)
\end{equation}
Where $\mathbf{m}_a(t)$ is the measurement of the atmosphere perturbation at instant $t$. We multiply both sides by 
$\mathbf{c}_{\xi}(t)^{T}$ and take the time average:
\begin{equation}
\label{eq:discreteopen2}
< \mathbf{m}_{\xi}(t) \cdot \mathbf{c}_{\xi}(t)^{T} > + < \mathbf{m}_a(t) \cdot \mathbf{c}_{\xi}(t)^{T} > 
= D < \mathbf{c}_{\xi}(t) \cdot \mathbf{c}_{\xi}(t)^{T} >
\end{equation}

\begin{figure*}
     \includegraphics[width=17cm]{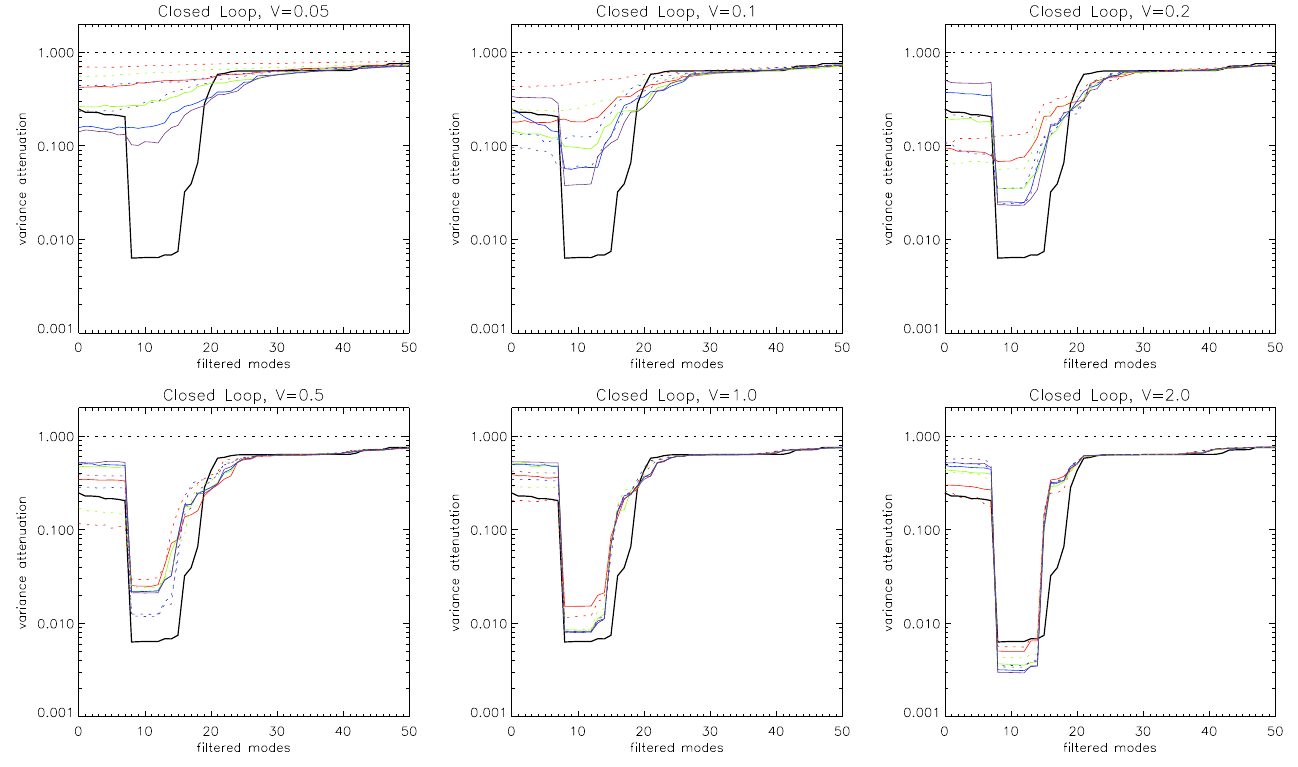}
    \caption{Variance attenuation $\Upsilon_{\sigma_{\epsilon}^2}$ as a function of filtered number of modes $\Xi$ for 
    interaction matrices obtained in closed loop with different levels of modulation on the DO-CRIME random vector; black is the
    poke matrix, red is 4000 iterations, green is 8000, blue 16000 and purple is 32000 iterations; full line is for $r_0$=20cm, 
    dashed line for $r_0$=10cm.  The larger the modulation, the better (lower) the attenuation, but at the expense of decreasing 
    Strehl ratio at the focal plane PSF (see Fig.~\ref{fig:strehl_cmatcl_min}).}
    \label{fig:covepsi_cmatcl_min}
\end{figure*}

We measure $\mathbf{m} = \mathbf{m}_{\xi} + \mathbf{m}_a$ but on average there is no correlation between 
the atmosphere and our noise signal $\mathbf{c}_{\xi}(t)$, such that
 $ < \mathbf{m}_a(t) \cdot \mathbf{c}_{\xi}(t)^{T} >  \rightarrow 0 $ and the interaction matrix is then given by:
\begin{equation}
\label{eq:intermatopen}
D = < \mathbf{m}_{\xi} \cdot    \mathbf{c}_{\xi}^T > \cdot <  \mathbf{c}_{\xi} \cdot  \mathbf{c}_{\xi}^T>^{-1}
\end{equation}
In practice we can use the measurement $\mathbf{m}$ in equation~\ref{eq:intermatopen} but we have
to obtain sufficient number of realisations for the covariance of the atmosphere and the random commands to 
become small enough as to become negligible. 
However we can improve our estimate of $\mathbf{m}_{\xi}$ by frequentially separating 
and rejecting the contribution of the atmosphere $\mathbf{m}_{a}$, 
which will be strongest at low temporal frequencies in the measurements $\mathbf{m}$ by 
restricting our random signal to high frequencies, which we can achieve by applying a high 
pass filter to both $\mathbf{c}_{\xi}$ and $\mathbf{m}_{\xi}$.
Therefore we set the amplitude of the random commands to  be sufficiently large to produce 
a meaningful signal on the wavefront sensor, though it may still be smaller than the low frequency 
atmosphere disturbances: it can be fine-tuned depending on the application (e.g. real time acquisition 
during science exposure, e.g. see Section~\ref{sec:closedloopsimul} and Fig.~\ref{fig:covepsi_cmatcl_min}). 
The reason for using a random command vector is that the the covariance 
matrix $<  \mathbf{c}_{\xi}(t) \cdot  \mathbf{c}_{\xi}^{T}(t) >$ (and for that matter
 $<  \Delta \mathbf{c}_{\xi}(t) \cdot  \Delta \mathbf{c}_{\xi}^{T}(t) >$ in the closed loop case) will be diagonal and thus well 
conditioned for inversion.

If there are invisible or poorly sensed modes in the system, they can be filtered during the SVD inversion process. However, 
if the control matrix $A$ is obtained directly from equation~\ref{eq:commat}, the measurements covariance 
matrix $< \mathbf{m}(t) \cdot  \mathbf{m}^{T}(t)  >$ will not be well conditioned and will need filtering at inversion, 
achieving the same effect. We also note in passing that we tried to use slope Hadamard matrices for exciting the
sensor response optimally as described by \cite{ref:meimon15}, but found no improvement in final accuracy with 
respect to using random vectors though our method uses more iterations due to the presence of turbulence 
that needs to be averaged out. We interpret this as whatever optimisation is gained in the 
Hadamard scheme is lost in the inversion of the commands (or measurements) covariance matrix inversion. 
The main reason why this method is robust is because
the $<  \mathbf{c}_t \cdot  \mathbf{c}_t^T>$ matrices are diagonal and can be inverted exactly.

\begin{figure*}
	\includegraphics[width=17cm]{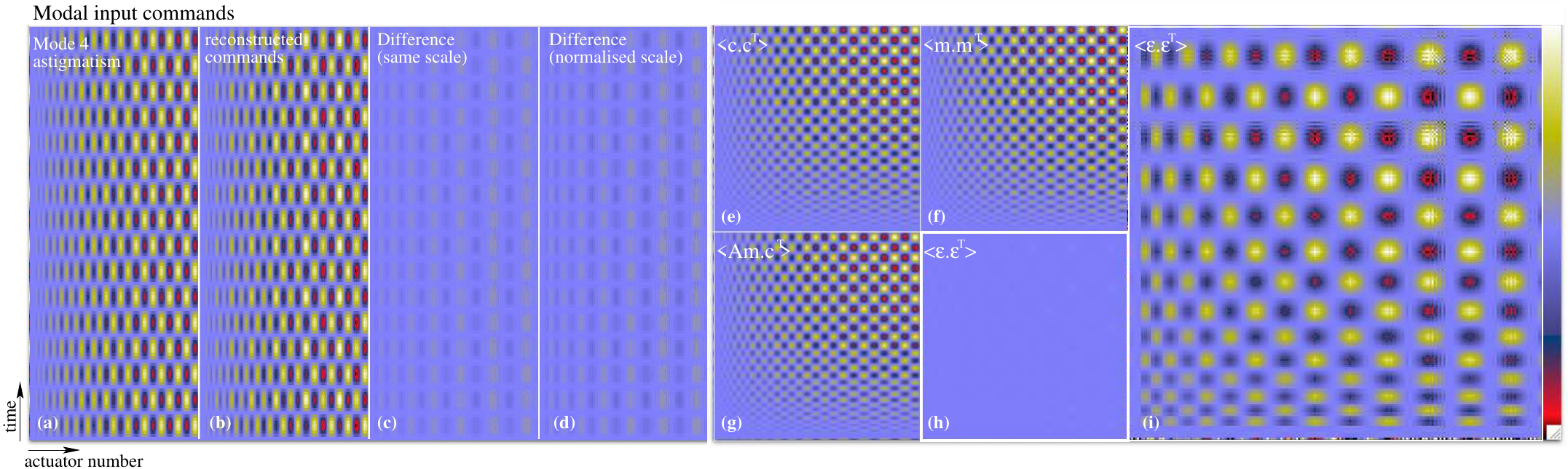}
    \caption{Modal (right) demonstration of the control matrix reconstruction fidelity simulation tool using mode 4, astigmatism. The panels are the
    same as in Fig.~\ref{fig:modulationsim1}. Panels (e), (f), (g) and (h) show the four terms of equation~\ref{eq:covepsilon}.} 
    \label{fig:modulationsim2}
\end{figure*}

\section{Simulations}
To test the DO-CRIME method we developed a simple simulation code based on a Shack Hartmann WFS model 
and the finite element model of the ASM influence functions; given any command vector $\mathbf{c}_{a}$, we can
compute the corresponding response on the wavefront sensor, and generate the $\mathbf{m}_{a}$ measurement vector. 
We used realistic values for the dimensioning of the system:
the `imaka ASM is expected to have 211 actuators and we used a $16 \times 16$ WFS to sample them. With this tool,
not only were we able to generate poke and DO-CRIME interaction matrices, but we were also able to assess the quality
of the derived control matrices. This issue of how effective a control matrix actually is, contains two components: the first is the actual 
numerical conditioning of the matrix, meaning that a given matrix will propagate noise depending on its condition number. 
This is given by $\sigma_{\mathbf{c}}^{2}(\Xi) = \Sigma$ Trace$[D^{+^{T}}(\Xi) \cdot D^{+}(\Xi)]$. Physically,
this noise propagation coefficient is the variance of the noise on the reconstructed commands when the input vector 
$\mathbf{m}$ contains 1rad$^2$ of noise. The trivial case of this is that  the noise propagation can be reduced by filtering out 
more modes, at the expense of accurate wavefront reconstruction. This leads us to develop a second quality criterion, namely the 
ability of the control matrix to reconstruct a wavefront for a given input vector. For example, if we modulate a given actuator 
at a known temporal frequency, and we measure the wavefront sensor vector, which we then multiply by the command matrix, 
how different is the reconstructed command vector to the input, how accurately does a given control matrix allow to reconstruct 
a command vector (Fig.~\ref{fig:modulationsim1}, Fig.~\ref{fig:modulation32})? This will of course depend on the 
input commands and their 
cross-correlations. If the input commands generate a white noise command vector, then the noise propagation coefficient 
determines the quality of the matrix. But if the commands contain low spatial frequencies, which produce cross correlations 
across different areas of the pupil, different modes may need to be filtered; for example, if we modulate an astigmatism on the
deformable mirror, a modal command matrix may be more effective than a zonal one (Fig.~\ref{fig:modulationsim2}).  

\begin{figure*}
	\includegraphics[width=12.5cm]{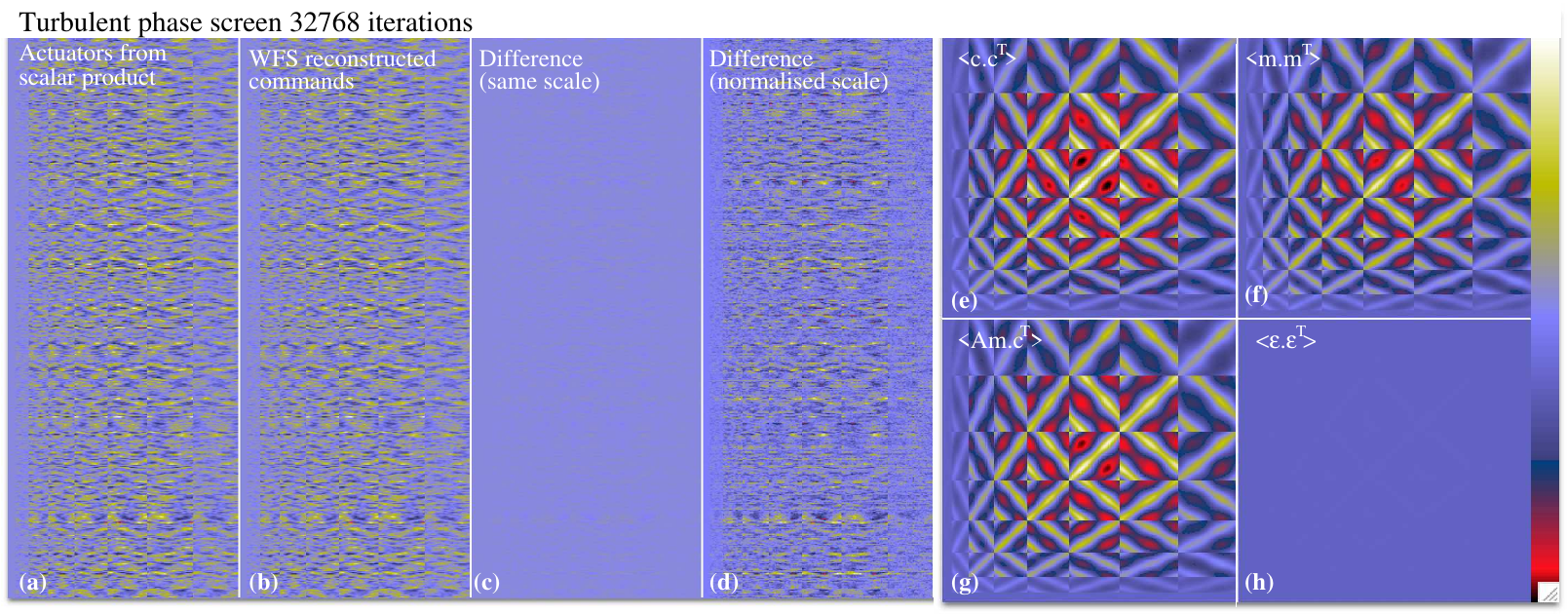}
	\includegraphics[width=5cm]{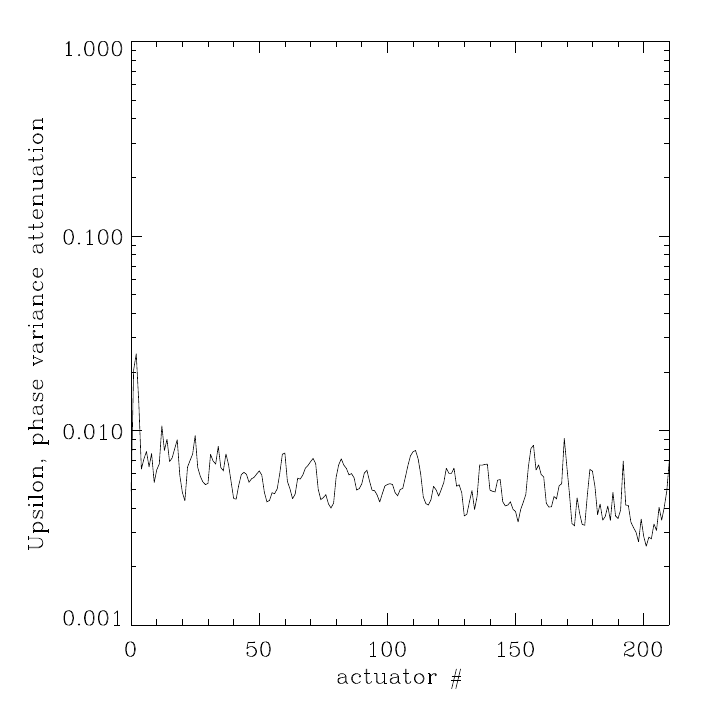}
    \caption{Same display as Fig.~\ref{fig:modulationsim1} but in phase space as opposed to actuator space, using 
    turbulent phase screens to generate the commands (by scalar product with the influence functions) and the measurement 
    through a model wavefront sensor (left). 
    The covariance matrices of the commands $<\mathbf{\varsigma} \cdot \mathbf{\varsigma}^T>$, of the measurements  
    $\Delta_{ij} D^+ < \mathbf{m} \cdot  \mathbf{m}^T> D^{+T} \Delta_{ij}^T$ 
    and of the cross terms $<\Delta_{ij} D^+  \mathbf{m} \cdot \mathbf{\varsigma}^T>$ are shown in the middle and are 
    similar in overall structure. However, their residual difference $<\mathbf{\epsilon} \cdot \mathbf{\epsilon}^T>$ (middle bottom 
    right) is crucial as it is a measure of the control matrix ability to reconstruct turbulent wavefronts. The ratio of the trace of 
    the residual phase covariance normalised to the input phase modulation, $\Upsilon_{\sigma_{\epsilon}^2}$ is shown 
    for $\Xi = 9$ on the right as a function of actuator number.} 
    \label{fig:modul_covmat_turbu}
\end{figure*}

We call this quality criterion of control matrices the control matrix {\em fidelity} and we can quantify it by comparing the 
known input commands $\mathbf{c}_a$ with the measured output $D^{+}(\Xi) \cdot \mathbf{m}_a$ and measuring the residual
difference $\mathbf{\epsilon}_a (\Xi)$:
\begin{equation}
\label{eq:epsilon}
\mathbf{\epsilon}_a(\Xi) = \mathbf{c}_{a} - D^{+}(\Xi) \cdot \mathbf{m}_{a}
\end{equation}
The optimal value of $\Xi_o$ (in other words, the optimal number of corrected modes) is the one that minimizes $\mathbf{\epsilon}_a$ for which we define the optimal control 
matrix $A = D^{+}(\Xi_o)$ We can write the covariance of the residual difference $<\mathbf{\epsilon}_a \cdot \mathbf{\epsilon}_a^T >$ as:
\begin{eqnarray}
\label{eq:covepsilon}
<\mathbf{\epsilon}_a \cdot \mathbf{\epsilon}_a^T > & = & < ( \mathbf{c}_{a} - D^{+} \cdot \mathbf{m}_{a} ) \cdot 
	( \mathbf{c}_{a} - D^{+} \cdot \mathbf{m}_{a})^T  > \nonumber \\
 & = & <\mathbf{c}_{a} \cdot \mathbf{c}_{a}^{T} > + D^{+} < \mathbf{m}_{a} \cdot \mathbf{m}_{a}^{T} > D^{+^{T}} - \nonumber \\
 & &  	 < \mathbf{c}_{a} \cdot \mathbf{m}_{a}^T D^{+^{T}}  + D^{+} \mathbf{m}_{a} \mathbf{c}_{a}^{T} > \nonumber \\
  & = & <\mathbf{c}_{a} \cdot \mathbf{c}_{a}^{T} > + D^{+} < \mathbf{m}_{a} \cdot \mathbf{m}_{a}^{T} > D^{+^{T}} - \nonumber \\
& &  	 < ( \mathbf{c}_{a} (D^{+} \mathbf{m}_{a})^{T}) + ( D^{+} \mathbf{m}_{a} \mathbf{c}_{a}^{T}  ) >
\end{eqnarray}
$<\mathbf{c}_{a} \cdot \mathbf{c}_{a}^{T} >$ is the statistical covariance matrix for which  
analytical expressions can be found in the literature \citep{ref:gendron94} and $< \mathbf{m}_{a} \cdot \mathbf{m}_{a}^{T} >$ is the slope 
covariance matrix for which it is also possible to compute analytical expressions based on the phase structure function \citep{ref:lai18}. 
We also note that that $\mathbf{c}_{a} (D^{+} \mathbf{m}_{a})^{T}  = ( D^{+} \mathbf{m}_{a} \mathbf{c}_{a}^{T}  )^{T}$, so the 
sum of the last two terms form a symmetric matrix. This equation tells us that the covariance of the residual of reconstruction 
are equal to the difference between the sum of the covariance of the commands and the covariance of the slopes (projected 
into command space) and the symmetric sum of their cross terms, as shown on panel (g) of Fig.~\ref{fig:modulationsim1},~\ref{fig:modulationsim2}. 
There are no analytical expressions for these cross terms, as they contain the interaction between measurements and commands, specific
to the adaptive optics system under consideration (i.e. actuator gains, misalignments, etc). 

Nevertheless we can compute these matrices numerically using our simulation tool: we generate vectors of commands
$\mathbf{c}_{a}$ according to the type of signal we want to reconstruct (in our case, Kolmogorov turbulence, but we can 
include noise or vibrations), and record their corresponding $\mathbf{m}_{a}$ responses. Then, for each control matrix 
of which we want to evaluate the fidelity, we compute $\mathbf{\epsilon}_a$ according to Equation~\ref{eq:epsilon} from 
which we compute the residual covariance matrix  $<\mathbf{\epsilon}_a \cdot \mathbf{\epsilon}_a^T >$. We can then 
compare sum of the traces of the different residual matrices associated with each control matrix. The commands 
$\mathbf{c}_{a}$, the reconstructed commands $\mathbf{m}_{a}$ and thus the residual $\mathbf{\epsilon}_a$ are in 
actuator space, and the variance attenuation shown in Fig.~\ref{fig:modulationsim1},~\ref{fig:modulationsim2} is therefore 
on the commands applied to the deformable mirror. However, we are more interested in the {\em phase} variance attenuation.
We can easily switch from actuator to phase space by multiplying $\mathbf{c}_{a}$ and $D^{+} \mathbf{m}_{a}$ by the geometrical 
covariance matrix \citep{ref:gaffard87}, defined as:
\begin{equation}
\Delta_{ij} =  \int_{\mathrm{pupil}} M_i(\mathbf{r}) . M_j(\mathbf{r}) d\mathbf{r}
\end{equation}
Where $M_i(\mathbf{r})$ and $M_i(\mathbf{r})$ are the $i$ and $j$ influence functions in units of phase over the pupil. 
In fact from our turbulence phase screen, we can obtain the phase coefficient $\mathbf{\varsigma}_a$ directly by computing 
$\int_{\mathrm{pupil}} \phi_{\mathrm{turb}} . M(\mathbf{r}) d\mathbf{r}$.
We can then define the phase variance attenuation $\Upsilon_{\sigma_{\epsilon}^2}(\Xi)$ as the ratio of the 
{\em phase} variance of $\mathbf{\epsilon}_a = \Delta_{ij}D^{+} \mathbf{m}_{a}$ to the input phase variance of the 
commands $\mathbf{\varsigma}_a$. 
 
 \begin{equation}
 \label{eq:attenuation}
 \Upsilon_{\sigma_{\epsilon}^2} (\Xi) = \frac{\sum \mathrm{Tr}( <\mathbf{\epsilon}_a \cdot \mathbf{\epsilon}_a^T > )}
 {\sum \mathrm{Tr}( <\mathbf{\varsigma}_a \cdot \mathbf{\varsigma}_a^T > ) }
 \end{equation}

\begin{figure*}
  \includegraphics[width=16cm]{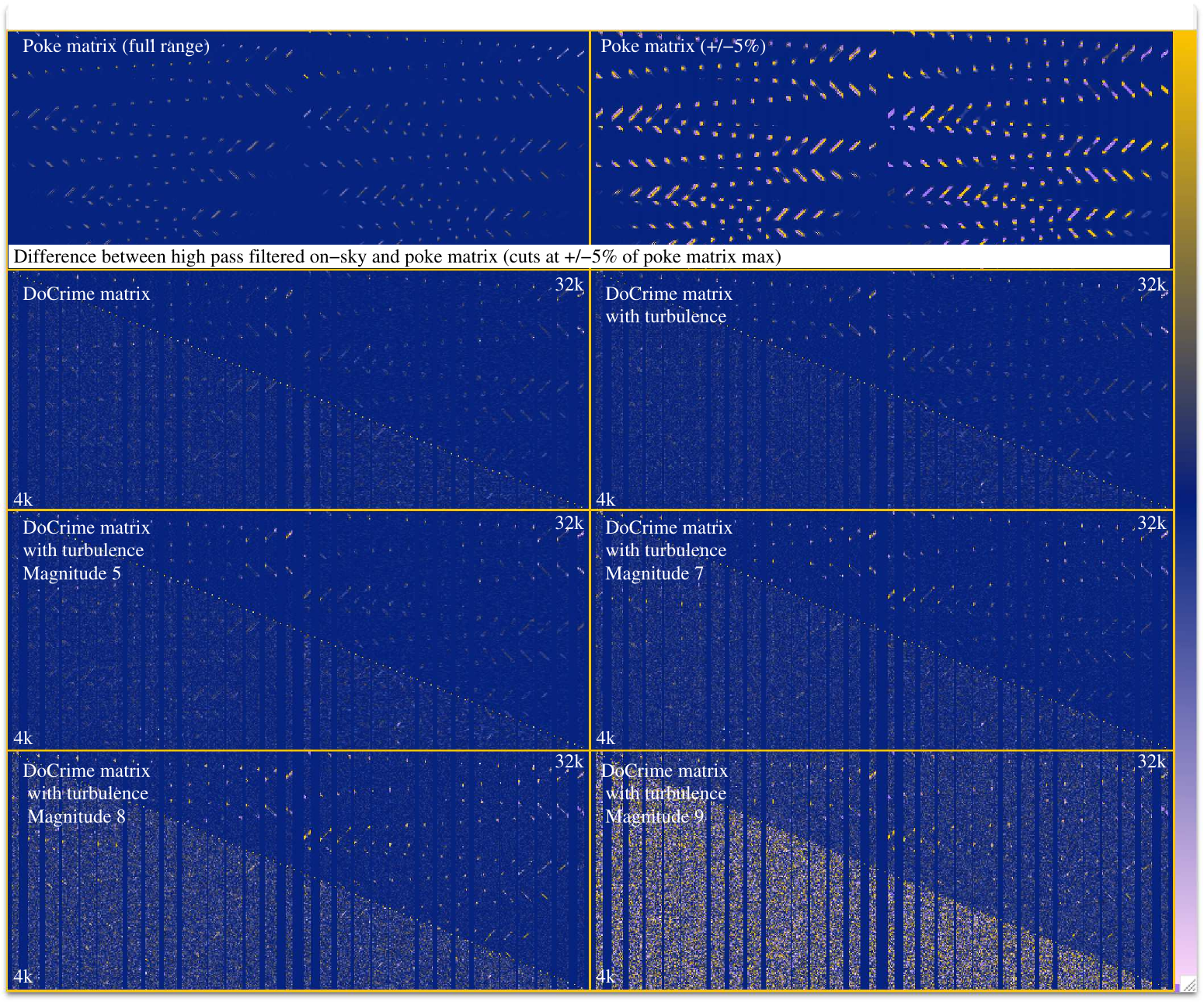}
    \caption{Top row, left: Poke interaction matrix, the shape of which depends on the geometry of the actuators and the subapertures.
    The radial pattern of actuators produces the osciallating pattern. Top right shows the same matrix with cuts 
    at $\pm5\%$ of the maximum. The lower rows show the difference between the poke matrices and the interaction matrices 
    obtained with the high pass filtered DO-CRIME method for the 6 different cases studied (no turbulence no noise, no noise but 
    turbulence, and turbulence with increasing levels of noise). We show the differences between the six cases and the poke 
    matrix for 4000 (bottom left half)and 32000 samples (top right half), with cuts at $\pm5\%$ of the maximum of poke matrix.}
    \label{fig:diffimats}
\end{figure*}

This phase variance attenuation is to be understood as the ability of the wavefront sensor and control matrix to 
reconstruct an input wavefront. As such it is a precise and robust prediction of open loop feed-forward adaptive optics. 
In closed loop however, the feedback correction introduces non-linear behaviour; some poorly conditioned modes 
can build up and make the loop unstable. We carried out extensive simulations and found that in closed loop, the optimal number 
of filtered modes can be different for the minimum phase variance attenuation  and the best Strehl ratio, 
but without any clear or obvious pattern (the optimal number of filtered modes in closed loop is always greater or equal to the
number of filtered modes for which $ \Upsilon_{\sigma_{\epsilon}^2}$ is minimum). Qualitatively we see that the feedback 
loop can slowly build up invisible or poorly sensed modes which are not part of the interaction matrix basis (for example the island effect 
modes, or partial waffles). This effect can be (and usually is) mitigated by implementing a leaky integrator which 
adds further non-linearities. However, apart from running Monte Carlo simulations with varying number of filtered 
modes (which is particularly computer intensive), we have not found a reliable way to predict the optimal number 
of filtered modes in closed loop. Nonetheless, $ \Upsilon_{\sigma_{\epsilon}^2}$ is a rigorous metric of wavefront
reconstruction fidelity and is a good proxy for closed loop performance.

\begin{figure*}
	\includegraphics[width=17cm]{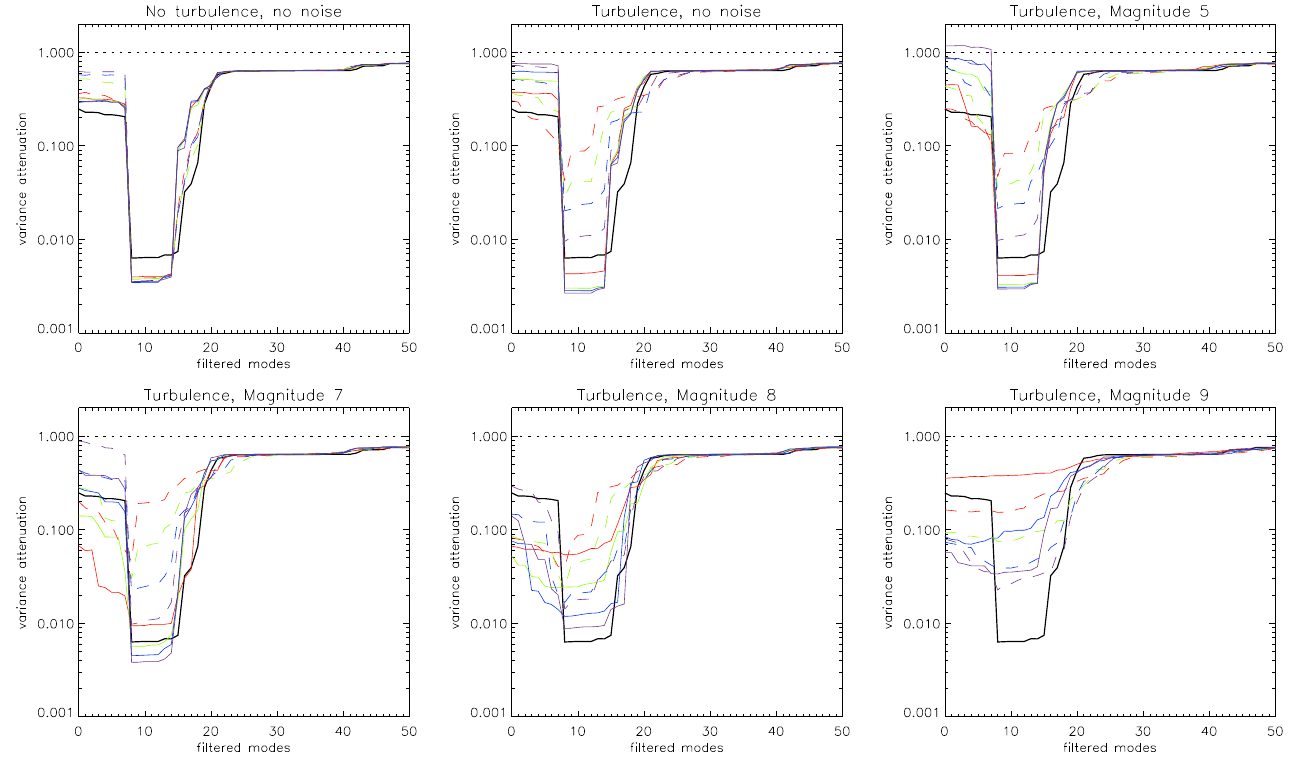}
    \caption{The attenuation of wavefront variance $\Upsilon_{\sigma_{\epsilon}^2}$ as a function of the number of filtered modes $\Xi$
    during SVD inversion.  When the number of filtered modes is too large, they asymptotically 
    approach unity as less of the wavefront is corrected. When too few modes are corrected, invisible modes are included at inversion 
    and make the loop unstable. The minimum (i.e. maximum attenuation) is around10 filtered modes (for this geometry). 
    Black is the poke matrix, red is the interaction matrix obtained with 4000 iterations, 
    green is with 8000, blue 16000 and purple with 32000 iterations (at 600 Hz, 7s, 14s, 28s, and 56s 
    respectively).  Full lines are high pass filtered data and full lines are the full data, containing the atmospheric signal. 
    It is clearly apparent that the presence of turbulence can be well mitigated 
    by the use of the high-pass filter, as can noise, as long as it doesn't dominate (up to magnitude 9, equivalent to 200 detected 
    photo-electrons per subaperture and per integration time, with 3$e^-$/pixel read noise and $12 \times 12$ pixels per subaperture)}
    \label{fig:covepsi_cmat}
\end{figure*}

To test out  the code we started with simple, synthetic signals: we applied zonal (actuating a few 
adjacent actuators with a sinusoidal modulation) to the ASM. We used a control matrix obtained from inverting a poke
 interaction matrix measured with the simulation tool. Sample results are displayed  on Fig.~\ref{fig:modulationsim1} where 
 3 actuators are modulated sinusoidally.  Modal modulation of the commands (modulating an astigmatism with a sinusoidal 
 excitation coefficient) is shown on Fig.~\ref{fig:modulationsim2}. In both cases, the control matrix is capable of 
 accurately reconstructing the input signal and the epsilon covariance matrices show little or no structure; this is especially 
 true in the case of the astigmatism, as the low order modes are better conditioned in control matrix (they have higher 
 eigenvalues in the interaction matrix). 
 
 To be able to assess the fidelity of the reconstruction for real systems in real world situations, we need to determine the 
 fidelity of control matrices to reconstruct atmospheric turbulence. We start with the open loop 
 case: we simulate a phase screen and project it onto the deformable mirror to get a command vector (by computing the scalar 
 product of the turbulent phase screen with the deformable mirror's influence functions as described previously). 
 Simultaneously, we record the associated wavefront sensor measurement. We then generate a sequence of 32000 
 commands and associated measurements by sliding the turbulent phase screen across 
 the pupil at are reasonable wind speed and $r_0$. We used the UH2.2m telescope entrance pupil with a $r_0$ of 0.2m, a 
 windspeed of 8m/s. These results are shown on Fig.~\ref{fig:modul_covmat_turbu}. Note that this figure is now in phase 
 space, unlike Figs.~\ref{fig:modulationsim1},~\ref{fig:modulationsim2} which are in command space. 
 We can see that the covariance matrices 
 of the commands $<\mathbf{\varsigma} \cdot \mathbf{\varsigma}^T>$, of the measurements  $\Delta_{ij}
 D^+ < \mathbf{m} \cdot  \mathbf{m}^T> D^{+T} \Delta_{ij}^T$ and 
 of the cross terms $<\Delta_{ij} D^+ \mathbf{m} \cdot \mathbf{c}^T>$ are quite similar in overall structure, but their residual 
 difference $<\mathbf{\epsilon} \cdot \mathbf{\epsilon}^T>$ is not zero and is a measure of the control matrix ability 
 to reconstruct turbulent wavefronts. On this figure we also plot $\Upsilon_{\sigma_{\epsilon}^2}$ for $\Xi = 9$ as a function of 
 actuator number: the central actuators are not as well attenuated because they 
 are partially hidden by the central obstruction; all other actuators attenuate the incoming phase 
 variance by a factor $\sim 5 \times 10^{-3}$. 
 
As our simulation tool now includes a sliding phase screen, we can also use it to generate DO-CRIME matrices in 
realistic on-sky conditions, including atmospheric turbulence: we compute the shape of the deformable mirror associated 
with our random command coefficient vector $\mathbf{c}_{\xi}(t)$ and add it to the turbulence phase screen 
from which we generate a 
measurement vector (to which we can add noise or other realistic disturbances such as vibrations) from which we 
can generate interaction matrices according 
to Equation~\ref{eq:intermatopen}. We can now study and optimize the input modulation for the DO-CRIME 
method using these simulations, first in open loop, and then in closed loop to determine how many samples and what 
amplitude of modulation are needed so that these on-sky matrices can still be accurately measured in the presence of
turbulence and noise.

\subsection{Open loop}
In open loop, we introduce a turbulent phase screen as well as noise on the wavefront sensor to study the control matrix 
fidelity as a function of number of iterations. With a sampling frequency set at 600Hz, sample sizes of 4096, 8192, 16384 and
32786, correspond to 6.8s, 13.7s, 27.3 and 55s respectively. We use a zero point of $9.6\times 10^{10} $ detected 
photons/second over the entire pupil for a magnitude zero star, with stars of magnitudes 5, 7, 8 and 9. In more useful units, 
these correspond to  3300, 500, 200 and 80 detected photo-electrons per subaperture and per integration time. We also use 
$3 e^{-}$/pixel read noise and $12 \times 12$ pixels per subapterture. We then generate interaction matrices using 
the random $\mathbf{c}_{\xi}$ command vectors and the associated measurements $\mathbf{m}_{\xi}$ for the 6 following 
cases: No turbulence, no noise; turbulence (phase screen with $r_{0}=0.2$m) but no noise; and turbulence with the 4 flux 
levels described above. We compute interaction matrices using the raw data as well as applying a high pass filter to better 
separate the turbulence from the random DO-CRIME signal, and display the 
results in Fig.~\ref{fig:diffimats}. Qualitatively, the interaction matrices generated in almost all 
cases with 32000 samples are indistinguishable from the poke interaction matrix (apart from the highest noise case), and even with 
4000 samples, the introduction of turbulence but no noise (or low noise level, magnitude 5) does not visually degrade the 
interaction matrix compared to the poke matrix.

To address this quantitatively, we invert these interaction matrices using Singular Value Decomposition, filtering from zero 
to 210 modes, and using our precomputed $\mathbf{\varsigma}_a$ and $\mathbf{m}_a$ vectors, we compute the control matrix fidelity 
$\Upsilon_{\sigma_{\epsilon}^2}$ as a function of $\Xi$, which we plot for the 6 cases and the 4 number 
of iterations in Fig.~\ref{fig:covepsi_cmat}. 

 \begin{figure}
	\includegraphics[width=8.5cm]{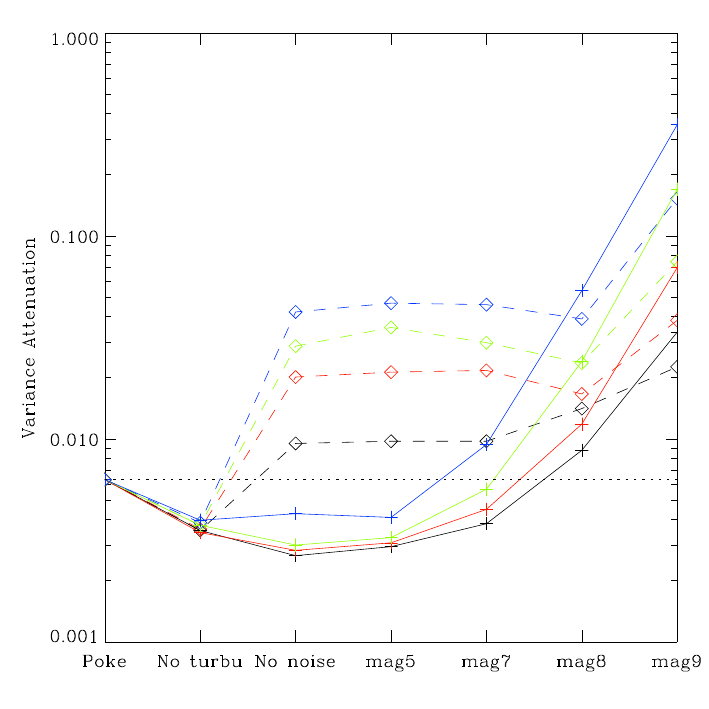}
    \caption{Minimum values of  $\Upsilon_{\sigma_{\epsilon}^2}$ for each case, obtained by filtering the optimal 
    number of modes $\Xi_o$ on Fig.~\ref{fig:covepsi_cmat}. Form left to right: Poke matrix; no turbulence and no noise; turbulence 
    and no noise, turbulence and magnitudes 5, 9, 10 and 11. Blue is the control matrix obtained with 4000 iterations, green is 
    with 8000, red 16000 and black with 32000 iterations, dashed lines are full data and full lines are high pass filtered.}
    \label{fig:covepsi_cmat_min}
\end{figure}

 We can point out a few salient features of these plots: The use of a high pass filter on the measurements allows to 
 efficiently remove the contribution of the turbulence which is mainly at low frequency. If noise is negligible, then  
 sequences with more than 8000 iterations (13 seconds @ 600Hz) are sufficient to measure interaction matrices 
 that perform just as well as poke matrices, i.e. same minimum value of $\Upsilon_{\sigma_{\epsilon}^2}(\Xi_o)$. 
 When the level of noise becomes more important, the fidelity of the control matrices can degrade 
 substantially, and increasing the number of samples or iterations doesn't seem to be improve the resulting interaction matrices. This
 is most likely due to the inclusion of read noise and no thresholding in the centroid computation, as one would expect the results 
 to degrade more gradually for photon noise only. 
 Also, the fact that the noise is uncorrelated from one frame to the next also means its covariance matrix is diagonal and 
 indistinguishable for the random commands sent on the DM, and high pass filtering cannot separate these contributions either. 
 Thus the method works best with stars that are sufficiently bright, using a high pass filter on the data to remove most of the contribution of the turbulence, and to increase the amplitude of the random modulation the a level such that its random fluctuations are larger than the 
 noise and thus dominate the signal on the wavefront sensor.
  
 \begin{figure}
         \includegraphics[width=8.5cm]{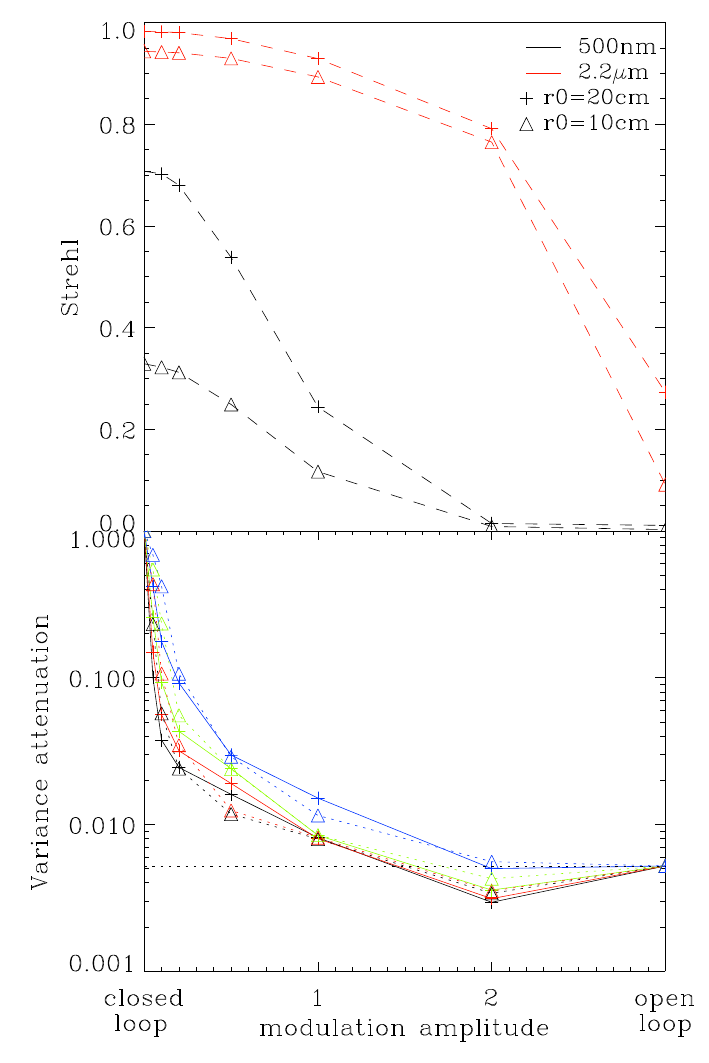}
\caption{Top: Strehl ratio attenuation (black: 500nm, red: 2.2$\mu$m) due to modulation on DM; bottom, minimum values of 
    $\Upsilon_{\sigma_{\epsilon}^2}$  for optimally filtered number of modes at inversion (blue: 4000, green: 8000, red: 
    16000, black: 32000 iterations, full line: $r_0$=0.2m, dashed line: $r_0$=0.1m). Closed loop (no modulation) 
    with poke matrix is shown for modulation amplitude of zero, and open loop (no modulation, no correction) is shown for 
    modulation amplitude of 3. The poke matrix variance attenuation is shown as the black dashed line.}
    \label{fig:strehl_cmatcl_min}
    \end{figure}
    
These results are summed up in Fig.~\ref{fig:covepsi_cmat_min}, which shows the minimum value of 
$\Upsilon_{\sigma_{\epsilon}^2}(\Xi_o)$ obtained by filtering the optimal number of modes on Fig.~\ref{fig:covepsi_cmat}.

When no turbulence and no noise is present, our matrices perform slightly better than the poke matrix because they 
naturally assign the optimal weight to partially illuminated edge subapertures (which is not done on the poke matrix), 
and 4000 iterations are sufficient to converge with or without high-pass filtering. The dashed lines show the results without 
temporal filtering and when turbulence is present, they do not perform
as well as the poke matrix; for 4000, 8000, 16000 and 32000 iterations, the phase variance attenuation goes from $4 \times 10^{-2}$ to
$3 \times 10^{-2}$ to $2 \times 10^{-2}$ to $1 \times 10^{-2}$  respectively. From this we can infer that 64000 iterations may provide
a comparable level of performance to the poke matrix but this is prohibitive to simulate and in fact not very useful on sky, as we can get 
much better results with a simple high pass filter on the data (full lines). The high pass filtered method works well in the presence of
turbulence and noise, up to magnitude 7 (500ph/subaperture/integration time) when the number of sample is $>$8000, but 
$\Upsilon_{\sigma_{\epsilon}^2}(\Xi_o)$ starts to degrade for magnitude 8 (200ph/subaperture/integration time) and higher.

\subsection{Closed loop simulations}
\label{sec:closedloopsimul}
Since our simulation tool allows to record the WFS measurements at each time step, it is  straightforward to implement a 
closed loop feedback using a simple integrator: we compute the shape of the deformable mirror based on the previous 
measurements (multiplied by a loop gain), which we add to the previous commands, and subtract this shape to our turbulent
phase screen. This allows us to correct the turbulent phase, while at the same time 
adding our random command vector $\mathbf{c}_{\xi}$. This closed loop approach, described in Equation~\ref{eq:intmat}, 
effectively applies a high pass filter by using the differential commands $\Delta \mathbf{c}_t = \mathbf{c}_{t+1} - 
\mathbf{c}_t$ (of which the turbulence component will be small, since turbulence evolves more 
slowly than the sampling frequency of the AO system). However, modulating the deformable mirror will 
inevitably degrade the PSF at the focal plane. We therefore wanted to determine the acceptable level of modulation to
still generate valid control matrices and the impact of this modulation level on the Strehl ratio of the delivered PSF. 
The motivation was to determine whether the DO-CRIME process could be permanently running in the 
background during observations, providing constant feedback to misalignments and updating optimal control matrices. 

To keep things manageable, we decided to neglect noise in these simulations, where we kept the same AO 
configuration, using the 211 actuator ASM on the UH2,2m telescope with a 16x16 SH WFS operating at 600Hz. 
We first ran the simulation with no feedback and no modulation, but recorded 
the long exposure PSF: the delivered Strehl ratio on the PSF  in open loop was 1.1\% and 0.2\% at 500nm and
27.2\% and 9.1\% at 2.2$\mu$m for $r_0$(500nm)=0.2m and 0.1m respectively. Closing the feedback loop with a gain of 0.6
(and a control matrix obtained by inverting a poke matrix), 
the corrected PSF Strehl ratio was 70.7\% and 32.9\% at 500nm and 98.2\% and 94.3\% at 2.2$\mu$m for $r_0$(500nm)= 
0.2m and 0.1m respectively. Note that these high values are simply due to the very high order system on a 2.2m telescope
for which $D/r_0 \sim 2$ at K band.

We tested modulations of the random command vectors with amplitudes of  0.05, 0.1, 
0.2, 0.5, 1.0 and 2.0V; the corresponding PSFs had Strehl ratios shown in Fig.~\ref{fig:strehl_cmatcl_min} (top). 
The results of variance attenuation $\Upsilon_{\sigma_{\epsilon}^2}$  are shown on the bottom panel of 
Fig.~\ref{fig:strehl_cmatcl_min}. As expected, as the modulation amplitude V becomes larger, the fidelity becomes 
better and just like in the open loop case, there is an improvement in going from 4000 to 8000 steps, but the performance
does not seem to improve by further increasing the number of steps to 16000 or 32000. Also not completely 
unexpected, if the modulation is negligible on the Strehl (values of the modulation of 0.05 -- 0.5V), the control matrix is 
degraded with respect to the poke matrix. But as the control matrices improve (modulation range of $1 \sim 2$V), 
it is at the expense of a non-negligible drop of the Strehl ratio in the corrected PSF. In fact a modulation of 2V
 is required to match the poke matrix performance, for which the Strehl ratio is 1.4\% and 0.9\% at 500nm and 79.1\% 
 and 76.5\% at 2.2$\mu$m for $r_0$=0.2 and 0.1m respectively. 
 
 This is allows to conclude that  while it is possible to continually monitor interaction matrices using the DO-CRIME method
 in closed loop in the background in GLAO mode in the near infrared, the drop in performance is not 
 really compatible with SCAO especially in the visible or in ExAO applications. In this case it may in fact be  preferable to 
 measure the interaction matrices between science exposures during camera readouts or filter switches, which is possible 
 because the process is quite fast (8000 iterations@600Hz, $\sim$13 seconds) and does not require any modification 
 in setup or hardware.
 
 \section{On-sky validation with the `imaka GLAO demonstrator}
 \subsection{Imaka GLAO demonstrator}
\begin{figure*}
	\includegraphics[width=17cm]{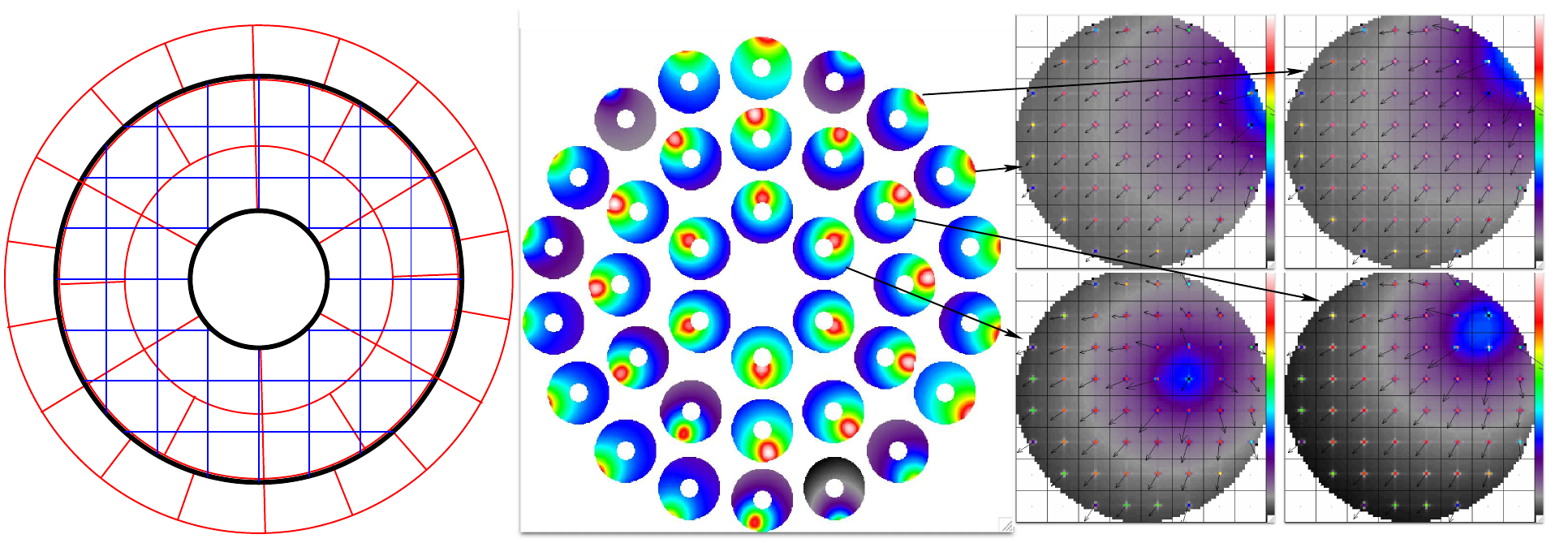}
    \caption{Left: the geometry of the $8 \times 8$ Shack-Hartman wavefront sensor (blue), the bimorph 
    mirror electrodes (red) and the pupil (black think lines). In the middle, the bimorph influence functions 
    are shown as a function of their position in the pupil. Right, the response on the Shack Hartman for electrodes 
    5, 16, 32 and 33 is shown, with Shack Hartman images overlaid to the phase and centroid vectors; note how 
    similar electrodes 32 and 33 (top row) appear to the sensor. The Shack Hartman spots are limited by the 
    diffraction of the subapertures.}
    \label{fig:imakageom}
\end{figure*}

`Imaka is a ground layer adaptive optics (GLAO) demonstrator currently deployed at the UH2,2m telescope \citep{lai08, chun18}. 
It uses five Shack-Hartmann wavefront sensors over a $24' \times 18'$ field of view, each with $8 \times 8$ 
subapertures, to control a 36 element bimorph mirror; a description of the performance of `imaka can be found in
\cite{abdu18}. The geometry of the `imaka system 
is shown on Fig.~\ref{fig:imakageom}.  An asymmetric Offner optical relay is used to re-image 
the pupil on the deformable mirror, but the system will soon be upgraded with an ASM developed by TNO to 
demonstrate the use of a new type of actuator based on the principle of magnetic reluctance. The use of an 
ASM on this Cassegrain telescope motivated the development of the DO-CRIME method, but the current optical 
relay provides an entrance focus which allows to confirm our proof-of-concept experimentally with this system by allowing a
direct comparison of DO-CRIME matrices and poke matrices, obtained using calibration sources. Unfortunately, `imaka 
interaction matrices are not ideal for this because in a hybrid system using Shack-Hartmann WFS with a bimorph mirror, 
outer ring adjacent electrodes generate almost identical measurement vectors, leading to poorly conditioned matrices and 
a certain level of degeneracy on the measurements for different actuators (see Fig.~\ref{fig:imakageom}, right).

\subsection{Laboratory tests using a single electrode}

\begin{figure*}
	\includegraphics[width=17cm]{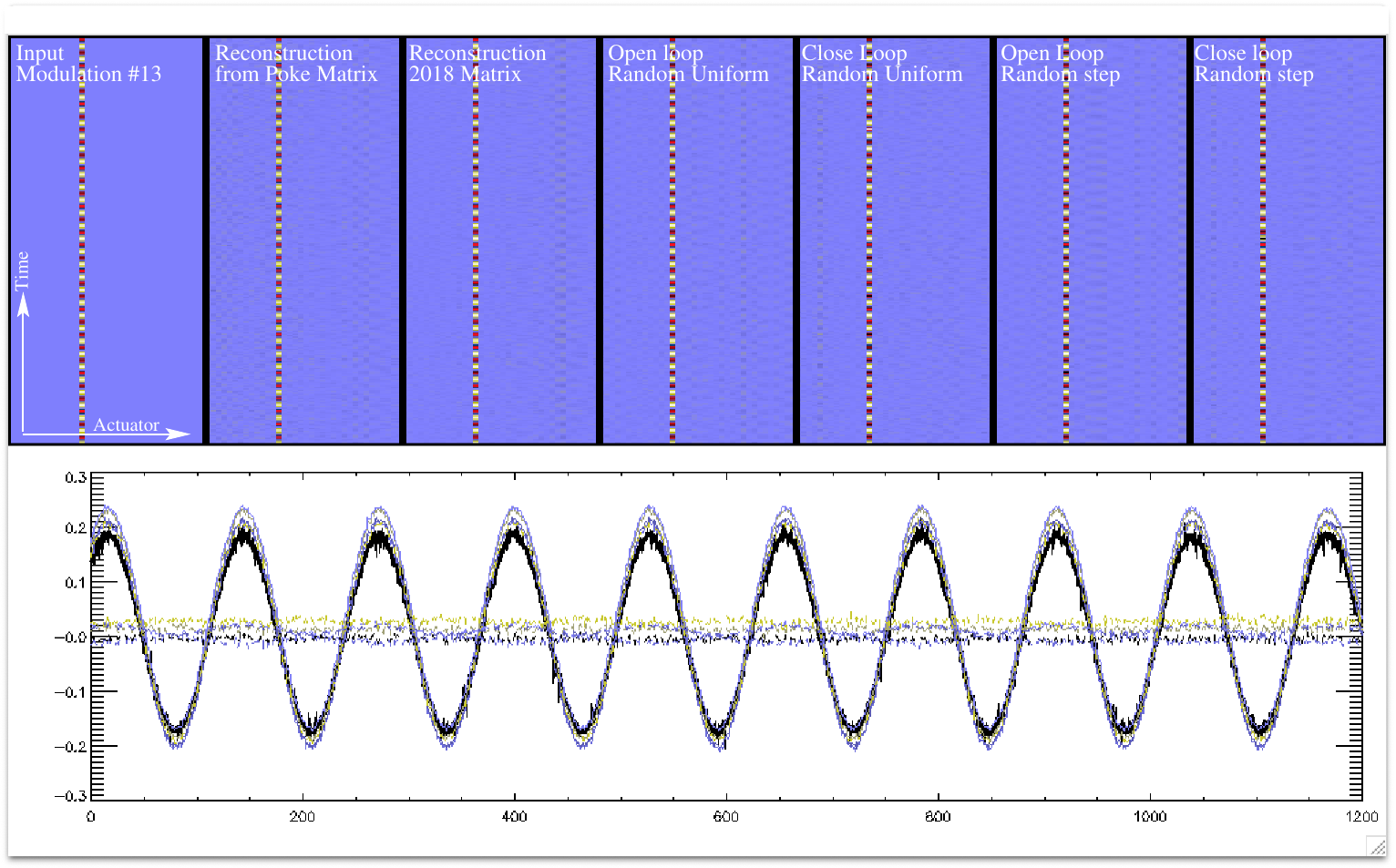}
    \caption{Known signal reconstruction through different control matrices: Electrode 13 is modulated with 
    a 32 time step period (top left panel), while all the other actuators are set to zero and measurements 
    are recorded on the wavefront sensor. The next 6 top panels show the  reconstructed commands obtained by
    $D.\mathbf{m}$, from left to right: poke matrix, old poke matrix (from 2018, showing that the system evolved) 
    and four different DO-CRIME matrices, open loop using a uniform random distribution, closed loop random 
    distrubution, open loop random step functions and closed loop random step functions. for each of the 
    DO-CRIME matrices, 13 modes had to be filtered at inversion to minimze the cross talk, compared to 7 
    filtered modes for the poke matrices. The bottom panel shows electrode 13's reconstructed modulation, 
    as well as electrode 32 (outer ring), to give an idea of the scale of the cross talk.}
    \label{fig:modulation32}
\end{figure*}

Before going on-sky where subtle differences in performance may be hidden by uncorrected free atmosphere
turbulence, we performed experiments with the `imaka AO system using artificial sources in a 
controlled laboratory setting. This means that the DO-CRIME matrices are obtained without the on-sky background turbulence, 
but the simulations have shown that we should be able to efficiently reject the atmospheric signal to extract the interaction matrices. 
The test signal $\mathbf{c}_a$ was the modulation of a single actuator (number 13) on the deformable mirror with a sine wave 
with a period of  32 time steps (at 180Hz) of amplitude 0.2V  for which we recorded the corresponding measurements 
$\mathbf{m}_a$ on the wavefront sensors. We reconstructed the command vector $D^{+} \mathbf{m}$ and compared it to 
the input command vector, shown on Fig.~\ref{fig:modulation32}. Modulating a single electrode is obviously not representative 
of atmospheric turbulence, but it does allow to see directly the level of crosstalk between electrodes in the reconstruction of 
each control matrix. We computed the reconstructed commands using different control matrices:
\begin{itemize}
\item A poke matrix obtained that day, using the average of 25 measurements poked
 positively and negatively 3 times for each actuator; 7 modes were filtered at inversion.
\item A poke matrix obtained in a similar fashion in January 2018, one year prior.
\item An open loop DO-CRIME matrix using a uniform random distribution with commands between $\pm 0.2$V and 27 000
time steps. This required filtering 13 modes at inversion.
\item A DO-CRIME matrix obtained in closed loop with the same uniform random distribution command vector, also requiring
13 filtered modes at inversion.
\item A DO-CRIME matrix obtained with commands of $+0.2$ and $-0.2$V randomly applied on each actuator in open loop, 13 modes filtered.
\item And a DO-CRIME matrix obtained in closed loop with the same binary command vector and same number of filtered modes.
\end{itemize}
The modulation of electrode 13 is well reconstructed by all matrices, except the year old one, implying that the system evolved in
some way over the course of this time. The open loop, binary matrix seems to minimize the 
amount of cross-talk, probably because it maximizes the amount of signal in the $\Delta \mathbf{c}_{\xi}$ vector to compute the
interaction matrix. 

\begin{figure*}
	\includegraphics[width=17cm]{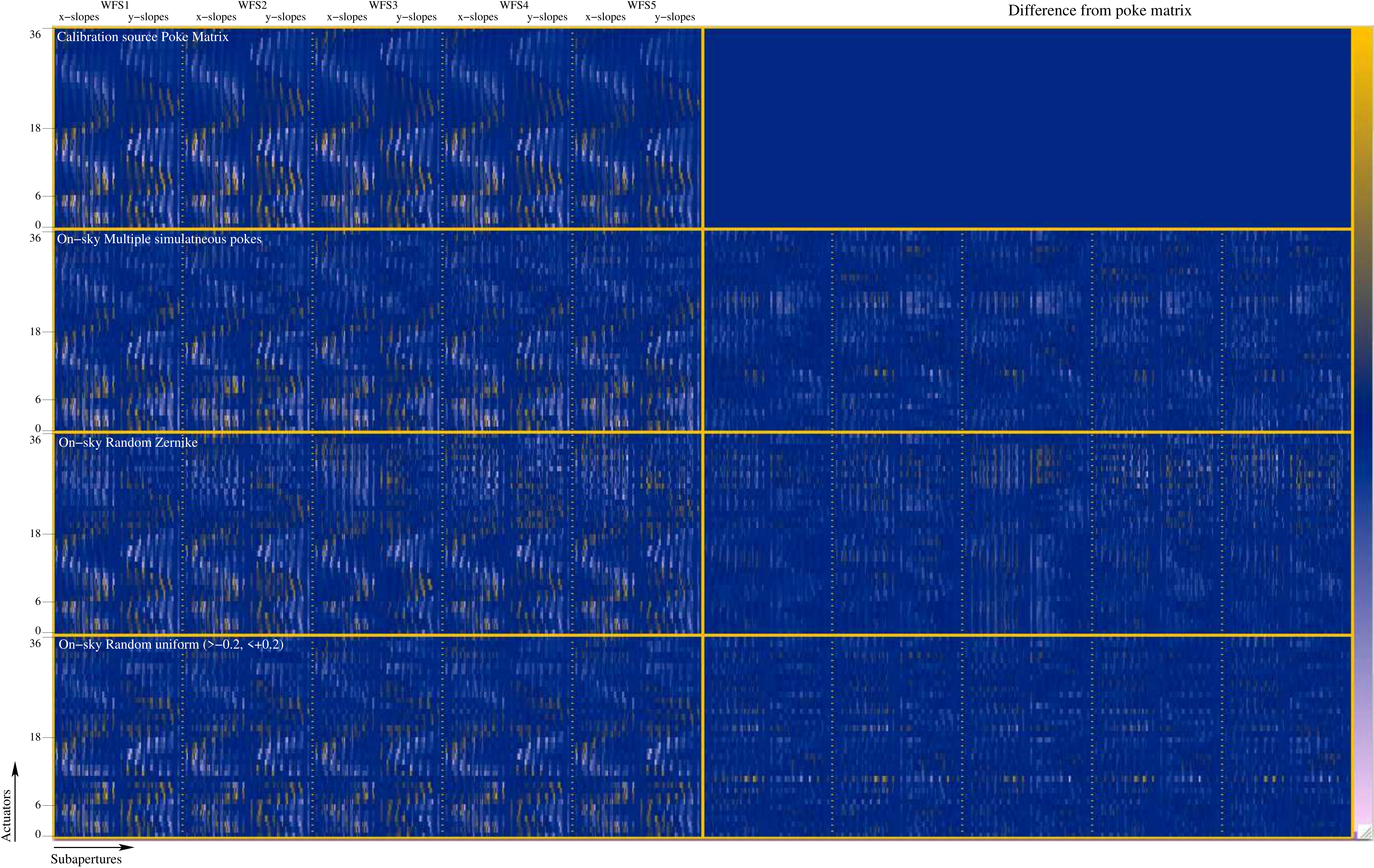}
   \caption{On-sky interaction matrices for imaka, using 5 SH wavefront sensors and a bimorph mirror. The 
   left column shows the matrices, while the right column shows the residuals between the left column 
   and the poke matrix obtained with the calibration source, which is shown on the top row. The second 
   row show shows an on-sky matrix obtained with simultaneous multiple (2 or 3 at each time step) 
    random poke commands. The next row shows the interaction matrix obtained on-sky using random Zernike 
    commands,  while the bottom row shows an interaction matrix obtained with a uniform 
    random distribution of commands between $-0.2$ and $+0.2$. Note how the measurements for the outer actuators are 
    more noisy for the Zernike case, due to poor conditioning of the commands covariance matrix. Also note the line in the 
    residuals at actuator 10 for the random distribution of commands matrices. See text for explanation.}
    \label{fig:onskymats}
\end{figure*}

\subsection{On-sky DO-CRIME matrices}

During an observing run  in January 2020, we obtained an interaction matrix 
using the poke method by  pushing and pulling on the actuators with a given voltage ($\pm 0.2$ 
out of a scale of $\pm 1$) multiple times and averaging the displacement of the 
wavefront sensor spots using a regular centroid algorithm. We then applied different sets of random command 
vectors $\mathbf{c}_{\xi}$ on sky using natural reference guide stars and recorded the wavefront sensor measurements in 
open, but also in closed loop (using the control matrix obtained from the poke interaction matrix). We implemented the
high pass temporal filter described above on the measurements (and the commands in closed loop), since most of the 
turbulence occurs at low temporal frequencies, and this improved the signal on the interaction matrices; we found no
difference between a Wiener filter (using open loop centroid measurements as input), and a sine squared high pass filter 
as long as the high temporal frequency content remained unaltered. The different types of random 
command vectors $\mathbf{c}_a$ that we tested on-sky included multiple simultaneous poke commands, random Zernike modes,  
and uniform random distributions of commands between $-0.2$ and $0.2$.  

We used these measurements to generate control matrices according to equations ~\ref{eq:commat},
~\ref{eq:intmat} and ~\ref{eq:svd}.  We show the interaction matrices, as well as the difference between the on-sky matrices 
and the poke matrix, on Fig.~\ref{fig:onskymats}. For clarity, we only display the interaction matrices obtained using 
equation~\ref{eq:intmat} in open loop, to avoid confusing or unconnected effects linked to matrix inversion or filtering.

These on-sky matrices were obtained 
with only 4096 samples and non negligible noise which partially explains why these matrices are noisier than the 
poke matrices. Nonetheless we found that the interaction matrices that showed the least amount of difference with the poke 
interaction matrix were the ones using random command vectors. The reason for this is that they provide 
the most diverse set of $\Delta \mathbf{c}$, and thus generate the most varied response on the WFS. 
As this matrix has a diagonal covariance matrices $<  \Delta \mathbf{c}_t \cdot  \Delta \mathbf{c}_t^T>$, it is well conditioned 
and the inversion is straightforward and accurate.  On the contrary, the commands covariance matrix for Zernike modes 
has non-diagonal terms, especially in the outer ring of actuators, which requires more filtering at inversion. 
This can be seen in the residuals on the third row of Fig.~\ref{fig:onskymats}. The multiple simultaneous poke commands simply did 
not provide enough separate occurrences of non-zero $\Delta \mathbf{c}$ to give a strong signal, visible espceically on the outer 
ring actuators (second row of Fig.~\ref{fig:onskymats}). Finally, the horizontal line at actuator 10 on the difference plot is an example of
the advantage of using the DO-CRIME method to incorporate the dynamic behavior of the AO system in the interaction 
matrices; we discuss this in the next section.

\subsection{Electrode 10}
\label{sec:electrode10}
\begin{figure}
	\includegraphics[width=8.5cm]{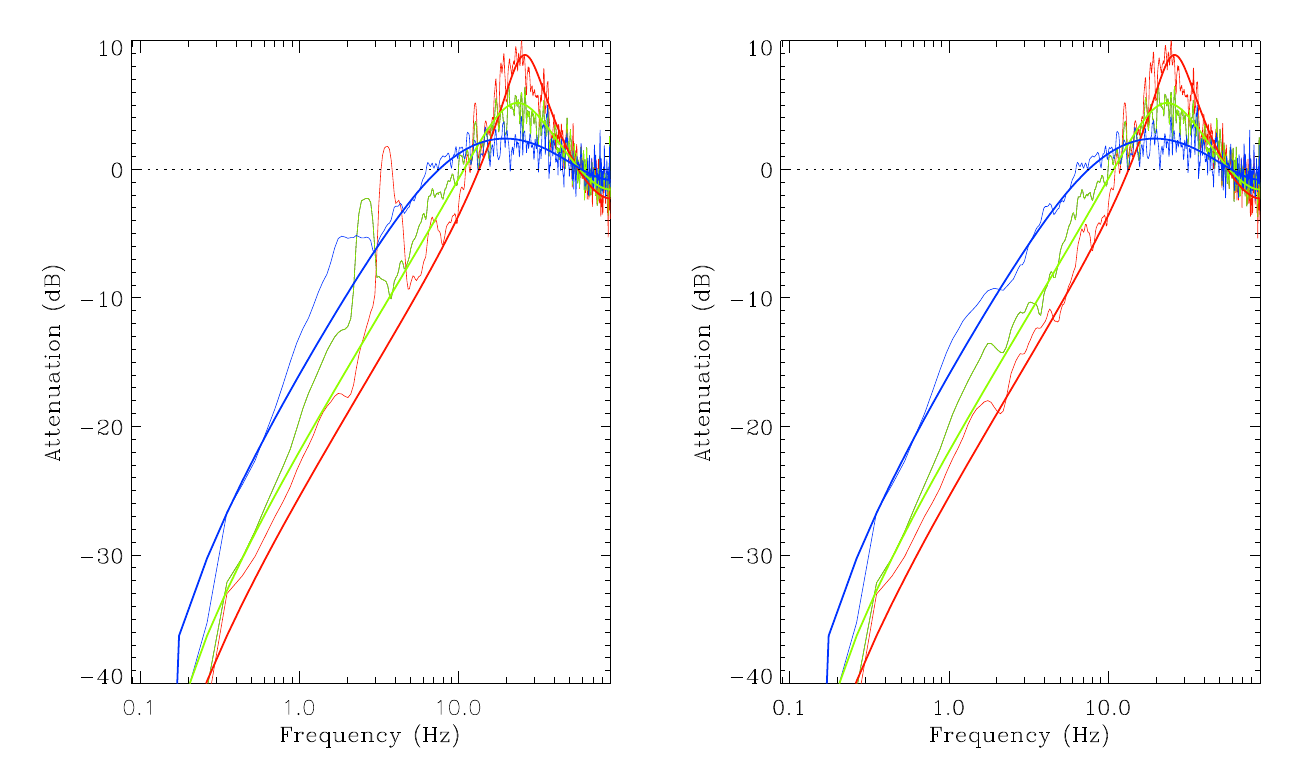}
    \caption{Error transfer functions for imaka with gains of 0.6 (red), 0.4 (green) and 0.1 (blue). The data is obtained in closed loop
    on an artificial source and no disturbance (the signal is photon noise). The thick line is a model based on the Fourier transform of
    a simple integrator discrete step function with pure delay of one frame. The left panel shows the error transfer function using all 
    the actuators and the right panel shows exactly the same data but removing actuator 10 from the analysis.}
    \label{fig:transfunc}
\end{figure}

While measuring the temporal transfer functions of imaka, the dynamical response of electrode 
number 10 was found to be  unreliable (Fig.~\ref{fig:transfunc}). The reason for this is currently 
under investigation but could be due to a poor connection of the 
electrode at the bimorph mirror, which could change its capacitance. However, because the DO-CRIME on-sky matrices are obtained 
dynamically, we see this row appear in the residuals column of Fig.~\ref{fig:onskymats}, as a marked difference 
between the static poke matrices and the on-sky matrices. At inversion, this row is automatically set to zero by SVD 
inversion and electrode 10 is thus filtered from the command matrix because its dynamic response is not sufficiently linear. 
In fact further laboratory testing showed that electrode 4 also has a slightly poorer dynamic response, although not at 
the same level as electrode 10, so it is usually left in the control matrix at inversion. This illustrates the advantage of 
measuring interaction matrices in the exact same conditions as those in which the system operates.

\subsection{On-sky validation}
\begin{figure*}
	\includegraphics[width=17.5cm]{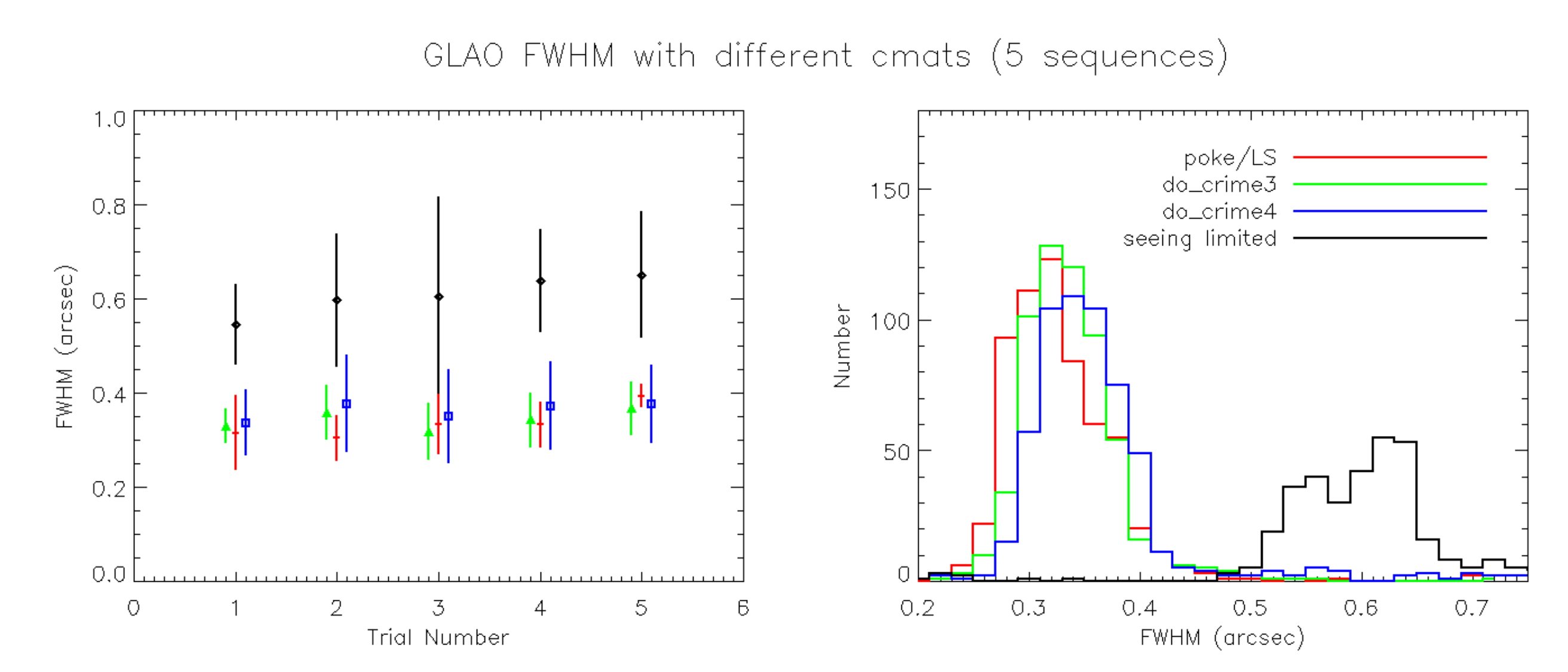}
    \caption{Average and standard deviation for 5 on-sky sequences (left) and histogram (right) of FWHM 
    obtained from on-sky focal plane images. Open loop data is shown in black; the double-peaked histogram suggests
    that the atmosphere was not stationary during these measurements. Poke matrices are shown red and the DO-CRIME 
    matrices (shown in green for DO-CRIME3 and blue for DO-CRIME4) provided an essentially equivalent level of
    performance: The DO-CRIME3 matrices sometimes marginally outperform the poke matrices 
    (trial 2 and 5), although the DO-CRIME4 matrices seem to have a systematically larger standard deviation. 
    Within error bars and intrinsic uncertainty due to seeing fluctuations, the DO-CRIME matrices provide a comparable 
    level of performance to the poke matrices.}
    \label{fig:onsky}
\end{figure*}

Finally we validated the proof of concept for on-sky matrices using the imaka GLAO system: we compared the FWHM
of focal plane images obtained with GLAO correction using poke matrices and two DO-CRIME matrices, 
obtained by combining the same sequence of 4096 random commands uniformly distributed between -0.2V and +0.2V on sky four times. 
We obtained five sequences where we cycled through open loop, poke matrices and so-called DO-CRIME 3 and DO-CRIME 4.
The results are shown on Table~\ref{tab:finalres} and in Fig.~\ref{fig:onsky}; because the residuals are dominated by the uncorrected free 
atmosphere, subtle reconstruction effects can easily be drowned out, and all matrices provide the same level of performance 
within error bars. During the 5  sequences, DO-CRIME 3 provided better performance than poke matrices (trial 2 and 5, Fig.~\ref{fig:onsky}, 
left), similar performance (trial 1 and 4) and slightly worse performance (trial  2), while DO-CRIME 4 seems to always have a larger
standard deviation in FWHM. It is thus very hard to distinguish the relative performance of these matrices, as those subtle differences 
can be caused by variations in seeing during our measurements: the open loop FWHM histogram is double peaked (and relatively 
broad $\sim0.1"$), implying that the atmosphere was not stationary with different regimes at different times and could easily have varied by 
more than 0.1" (mean) during our acquisition sequence.

It is unfortunate that the Shack-Hartman-bimorph hybrid configuration generates invisible modes and that the interaction matrices 
are poorly conditioned, as it makes the performance more sensitive to noise and turbulence and thus it is difficult to assess how 
effective the control matrices are. Nonetheless, the on-sky performance of DO-CRIME matrices obtained on-sky is comparable to that
of poke matrices, at least within error bars. This is comforting as it ensures that we will be able to control the planned ASM for 
our Cassegrain telescope once it is delivered.

\begin{table}
\centering
\caption{FWHM obtained from focal plane images.}
\label{tab:finalres}
\begin{tabular}{|l||c||c|}
FWHM & Median(") & stdev(") \\ \hline
Poke & 0.321 & 0.12 \\
docrime3 & 0.334 & 0.12 \\
docrime4 & 0.345 & 0.14 \\
Seeing & 0.593 & 0.25 \\
\end{tabular}
\end{table}

\section{Conclusions}
We have presented a novel method for measuring interaction matrices on sky, when no intermediate focus for a 
calibration source exists, such as with convex adaptive secondary mirrors. The method consists in modulating the deformable
mirror using a random (but known) sequence of commands and measuring the associated simultaneous wavefront 
sensor measurements. Using the property of the modulation covariance matrix being diagonal, thus trivial to invert, we can
obtain interaction matrices by multiplying the cross-covariance of the measurements and commands by the inverse of the
commands covariance matrix. We have tested the method by simulation using turbulent phase screens, finite element models
 of the 211 actuator ASM influence functions and a model of our wavefront sensor, both in open loop, where we investigated the
 impact of measurement noise, as well as the use of high pass filtering to better separate the modulation from the turbulent
 signal, and in closed loop. In this latter case, we recorded the delivered Strehl ratio at the PSF focal plane to determine an 
 acceptable level of modulation for background operation. We find that in the GLAO regime, modulation at levels much smaller
 than the residual phase error still allow us to generate acceptable interaction matrices. For SCAO and ExAO, this method 
 is incompatible with the requirements on the accuracy of the control matrix and the level of image degradation in the focal plane. 
 In that case it is therefore probably preferable to measure interaction matrix between science exposures. The need for a constantly 
 updated control matrix is not obvious for most science cases performance, and since adaptive optics systems operate in closed 
 loop, as long as unstable modes are filtered out, errors in the control matrix translate in slightly delayed responses; apart from 
 at visible wavelengths or in cases of exquisite PSF stability, most adaptive optics programs can tolerate slightly suboptimal 
 control matrices, which also explains why they are often neglected in practice. Our method differs slightly from other methods 
 found in the literature in that it is invasive, but very fast, unlike other invasive methods, but it is also completely model independent
 unlike non-invasive methods.
 
 Nonetheless, we have shown that there are advantages to measuring interaction matrices in the same exact conditions dynamic
 conditions than those in which they will be used. For example, our method allows us to confirm that electrode \#10 on the bimorph
 mirror was misbehaving due to a probable bad connection on its connector, slowing down its response. It still showed up in the poke interaction
 matrix, but was degrading performance when dynamically excited. The DO-CRIME method naturally filters it out, as well as applies
 a gain to other actuators which may also have non-optimal response. Other examples of benefits to measuring the interaction 
 matrices on sky include accurate treatment of vignetting and diffraction of edge subapertures (which will be filtered out if 
 they become non linear), signal to noise ratio of each subaperture included in the control matrix, centroid gain in the case of 
 quad cell sensors, etc.

We have demonstrated that our method works on sky using the imaka GLAO instrument at the UH2,2m telescope, by generating
control matrices using different types of modulation commands (uniform, stepwise, modal) and were able to close the GLAO loop
on matrices acquired on sky. This confirms that our method will allow us to control our convex adaptive secondary mirror 
as soon as it is installed on the telescope. Tools to interpret control matrices in terms of parametric behavior (centering, rotation, 
magnification, distortion, vignetting) have also been developed for the integration of the ASM, in order to take advantage of the capability 
of the DO-CRIME method to generate interaction matrices on the fly. In a first step, we will be able to use them as a tool for 
alignement on-sky, but more generally also as a diagnostic tool providing a physical interpretation of the AO system state.

\section*{Acknowledgements}

`imaka is supported by the National Science Foundation under Grant No. AST-1310706 and by the Mount Cuba Astronomical Foundation.   
We thank Stefan Kuiper (TNO) for providing the model influence functions for the UH88 adaptive secondary mirror.  The authors also wish to 
recognize and acknowledge the very significant cultural role and reverence that the summit of Maunakea has always had within the indigenous 
Hawaiian community. We are most fortunate to have the opportunity to conduct observations from this mountain.




\vskip6pt{\large\it UH:2.2m:}  (`imaka) 

\section*{Data availability statement}
The data and simulation code underlying this article will be shared on reasonable request to the corresponding author.




\begin{thebibliography}{99}

\bibitem[\protect\citeauthoryear{Abdurrahman et al.}{2018}]{abdu18}
Abdurrahman, F. N., Lu, J. R., Chun, M., Service, M. W., Lai, O., F\"{o}hring, D.,
 Toomey, D., Baranec, C., 2018, "Improved Image Quality over $10^{\circ}$ Fields with 
 the `Imaka Ground-layer Adaptive Optics Experiment". The Astronomical Journal, 
 156(3), id.100, 17pp.

\bibitem[\protect\citeauthoryear{Babcock}{1953}]{ref:babcock53}
Babcock H., 1953, "The Possibility of Compensating Astronomical Seeing". Publications of the 
Astronomical Society of the Pacific, 65(386), 229.

\bibitem[\protect\citeauthoryear{Bechet et al.}{2011}]{ref:bechet11}
Bechet, C., Kolb, J., Madec, P.-Y., Tallon, M., Thi\'ebaut, E., "Identification of system
misregistration during AO-corrected observations", Proc. AO4ELT2, 2011, 
http://ao4elt2.lesia.obspm.fr/sites/ao4elt2/IMG/pdf/056bechet.pdf

\bibitem[\protect\citeauthoryear{Bechet et al.}{2012}]{ref:bechet12}
Bechet, C., Tallon, M., Thi\'ebaut, E., "Optimization of adaptive optics correction during observations:
Algorithms and system parameters identification in closed loop", Proc. SPIE Adaptive Optics Systems III, 2012, 
Vol. 8447, 8447C-1.  

\bibitem[\protect\citeauthoryear{Chun et al.}{2018}]{chun18}
Chun, M., Lu, J., Lai, O., Abdurrahman, F., Service, M., Toomey, D., Fohring, D., Baranec, C., 
Hayano, Y., Oya, S., 2018, "On-sky results from the wide-field ground-layer adaptive optics 
demonstrator 'imaka". Proc. SPIE Adaptive Optical Systems, 2018. Vol. 10703, p. 7.

\bibitem[\protect\citeauthoryear{Esposito et al.}{2006}]{ref:esposito06}
Esposito, S., Tubbs,R. , Puglisi, A., Oberti, S., Tozzi, A., Xompero, M., Zanotti, D., “High SNR  
measurement of interaction matrix on-sky and in lab” Proc. SPIE Advances in Adaptive Optics II, 2006, Vol. 6272, 
62721C.  

\bibitem[\protect\citeauthoryear{Gaffard \& Boyer}{1987}]{ref:gaffard87}
Gaffard J.P., Boyer C., “Adaptive optics for optimization of image resolution”, Applied
Optics, Vol 26, p 3772, (1987).


\bibitem[\protect\citeauthoryear{Gendron \& L\'ena}{1994}]{ref:gendron94}
Gendron, E., L\'ena, P., "Astronomical adaptive optics. I. Modal control optimization", Astron. \&
Astrophys., 291, 337--347.

\bibitem[\protect\citeauthoryear{Heritier et al.}{2017}]{ref:heritier17}
Heritier, C.T., Fusco, T., Neichel, B., Esposito, S., Oberti, S., Correia, C., Sauvage, J.-F., 
Bond, C., Fauvarque, O., Pinna, E., Agapito, G., Puglisi, A., Kolb, J., Madec, P.-Y., Bechet, C.,
 "Overview of AO calibration strategies in the ELT context". Proc AO4ELT5, 2017,
 http://research.iac.es/congreso/AO4ELT5/media/proceedings/proceeding-035.pdf

\bibitem[\protect\citeauthoryear{Heritier et al.}{2018}]{ref:heritier18}
Heritier, C.T., Esposito, S., Fusco, T., Neichel, B.,Oberti, S., Briguglio, R., Agapito, G., 
Puglisi, A., Pinna, E., Madec, P.-Y., "A new calibration strategy for adaptive telescopes with 
pyramid WFS". Mon. Not. Royal Astron. Soc., 481(2), 2018, 2829-2840.

\bibitem[\protect\citeauthoryear{Heritier et al.}{2019}]{ref:heritier19}
Heritier, C.T., "Innovative Calibration Strategies for Large Adaptive Telescopes with 
Pyramid Wave-Front Sensors". Ph.D. thesis Aix Marseille Universit\'e, 2019, 
HAL Id : tel-02390861.

\bibitem[\protect\citeauthoryear{Kasper et al.}{2004}]{ref:kasper04}
Kasper, M., Fedrigo, E., Looze, D.P., Bonnet, H., Ivanescu, L., Oberti, S., "Fast calibration of high order
adaptive optics systems". J. Opt. Soc. Am. A, 21(6), 2004, 1004--1008.

\bibitem[\protect\citeauthoryear{Kolb et al.}{2012}]{ref:kolb12}
Kolb J., Madec, P.-Y., Le Louarn,  M., Muller, N., B\'{e}chet, C., "Calibration strategy of the  AOF," Proc. SPIE Adaptive 
Optics Systems III, 2012, Vol. 8447, 84472D. 

\bibitem[\protect\citeauthoryear{Meimon et al.}{2015}]{ref:meimon15}
Meimon, S., Petit, D., Fusco, T., "Optimized calibration strategy for high order adaptive optics 
systems in closed-loop: the slope-oriented Hadamard actuation," Optics Express, 23(21), 2015, 27134--27144. 

\bibitem[\protect\citeauthoryear{Lai et al.}{2000}]{ref:lai00}
Lai, O., Stomski, P., Gendron, E., "MANO: the modal analysis and noise optimization program for the 
W.M. Keck Observatory adaptive optics system", Proc. SPIE Adaptive Optical Systems Technology 
Vol. 4007, 2000, 620-631.

\bibitem[\protect\citeauthoryear{Lai et al.}{2008}]{lai08}
Lai, O., Chun, M., Cuillandre, J.-C., Carlberg, R., Richer, H., Andersen, D., . Pazder, J., 
Tonry, J., Doyon, R., Thibault, S., Dunlop, J., Pritchet, C., V\'{e}ran, J.P., Ftaclas, C., 
Onaka, P., Hodapp, K.W., McLaren, R.A.. Bertin, E., Mellier, Y., Astier, P., Pain, R., 2008, 
"IMAKA: imaging from Mauna KeA with an atmosphere corrected 1 square degree optical 
imager". Proc SPIE Adaptive Optics Systems, 2008, Vol. 7015, p.12.

\bibitem[\protect\citeauthoryear{Lai et al.}{2018}]{ref:lai18}
Lai, O., Chun, M., Abdurrahman, F., Lu, J., Service, M., Fohring, D.,Toomey, D.,
"Deconstructing turbulence and optimizing GLAO using imaka telemetry".  Proc. SPIE Adaptive 
Optical Systems, 2018. Vol. 10703, 107036D.

\bibitem[\protect\citeauthoryear{Neichel et al.}{2012}]{ref:neichel12}
Neichel B., Parisot, A., Petit,  C., Fusco, T., Rigaut, F., "Identification and calibration of the interaction matrix 
parameters for AO and MCAO systems", Proc. SPIE Advances in Adaptive Optics II, 2006, Vol. 6272, 627220. 

\bibitem[\protect\citeauthoryear{Oberti et al.}{2006}]{ref:oberti06}
Oberti, S., Quir\'os-Pacheco, F., Esposito, S., Muradore, R., Arsenault, R., Fedrigo, E., 
Kasper, M., Kolb, J., Marchetti, E., Riccardi, A., Soenke, C., Stroebele, S., "" Proc. SPIE Adaptive 
Optics Systems III, 2012, Vol. 8447, 84475N. 


\bibitem[\protect\citeauthoryear{Pieralli et al.}{2008}]{ref:pieralli08}
Pirealli, F., Puglisi, A., Quiros Pacheco, F., Esposito, S., "Sinusoidal calibration technique for 
Large Binocular Telescope system," Proc. SPIE Adaptive Optics Systems III, 2008, Vol. 7015, 70153A-7. 

\bibitem[\protect\citeauthoryear{Roddier}{1988}]{ref:roddier88}
Roddier, F., 1988, "Curvature sensing and compensation: a new concept in adaptive optics". 
Applied Optics 27(7), 1988, 1223-1225.

\bibitem[\protect\citeauthoryear{Roddier}{1999}]{ref:roddier99}
Roddier, F., 1999, "Adaptive Optics in Astronomy". Cambridge: Cambridge University Press.

\bibitem[\protect\citeauthoryear{Tyson}{1991}]{ref:tyson91}
Tyson, R.K., 1999, "Principles of Adaptive Optics". Cambridge: Academic Press.

\bibitem[\protect\citeauthoryear{Wildi \& Brusa.}{2004}]{ref:wildi04}
Wildi, F.P., Brusa, G., "Determining the interaction matrix using starlight", Proc. SPIE Advancements 
in Adaptive Optics, 2004, Vol. 5490, 164--173. 

\bibitem[\protect\citeauthoryear{Woillez et al.}{2019}]{ref:woillez19}
Woillez, J., Abad, J., Abuter, R., Carpentier, E., Alonso, J., et al., "NAOMI: the adaptive optics system
of the Auxiliary Telescopes of the VLTI. Astronomy and Astrophysics" A\&A, 629, A41 (2019) 

\end{thebibliography}



\bsp	
\label{lastpage}
\end{document}